\documentclass[a4paper,USenglish,cleveref,autoref,thm-restate]{lipics-v2021}

\hideLIPIcs

\pdfoutput=1

\newif\ifEditMode

\EditModetrue

\usepackage{aliases}
\usepackage{preamble}
\usepackage{svg}
\usepackage{pifont}
\title{Understanding Blockchain Governance: Analyzing Decentralized Voting to Amend DeFi Smart Contracts}

\titlerunning{Understanding Blockchain Governance}

\author{Johnnatan Messias}{MPI-SWS \and \url{https://johnnatan-messias.github.io}}{}{https://orcid.org/0000-0002-6021-8402}{}

\author{Vabuk Pahari}{MPI-SWS}{}{}{}

\author{Balakrishnan Chandrasekaran}{Vrije Universiteit Amsterdam}{}{}{}

\author{Krishna P. Gummadi}{MPI-SWS}{}{}{}

\author{Patrick Loiseau}{Inria, FairPlay team}{}{}{}

\authorrunning{J. Messias et al.} 

\keywords{DAOs, Decentralized Governance, Token Concentration, Fairness}

\ccsdesc[100]{ }

\acknowledgements{
\section*{Acknowledgments}\label{sec:ack}

This research was supported in part by a European Research Council (ERC) Advanced Grant ``Foundations for Fair Social Computing'', funded under the European Union's Horizon 2020 Framework Programme (grant agreement no. 789373). It has also been supported by MIAI @ Grenoble Alpes (ANR-19-P3IA-0003) and by the French National Research Agency under grant ANR-20-CE23-0007. \red{Paper last revised on \today}.

}

\nolinenumbers %

\begin{document}

\maketitle

\begin{abstract}
  
Decentralized Autonomous Organizations (DAOs) have emerged as a novel governance mechanism in blockchain ecosystems, particularly within Decentralized Finance (DeFi). By enabling token holders to propose and vote on protocol changes, these systems promise transparent and equitable decision-making without centralized control. In this paper, we present an in-depth empirical study of the governance protocols of Compound and Uniswap, two of the most widely used DAOs in DeFi. Analyzing over 370 governance proposals and millions of on-chain events from their inception until August 2024, we uncover significant centralization of voting power: \stress{as few as 3–5 voters were sufficient to sway the majority of proposals}. We also find that the cost of voting disproportionately burdens smaller token holders, and that strategic voting behaviors, such as delayed participation and coalition formation, further distort governance outcomes. Our findings suggest that despite their decentralized ideals, current DAO governance mechanisms fall short in practice.

\end{abstract}

\section{Introduction}\label{sec:introduction}

Decentralized Finance (DeFi) applications have radically transformed banking and financial systems by providing a trustless, peer-to-peer alternative to traditional financial services~\cite{adams2021uniswap,Daian@S&P20,Perez@FC21,Qin@FC21}. Their ability to facilitate interactions between participants without the need for intermediaries is underpinned by blockchain technology, where compliance is ensured through smart contracts~\cite{Nakamoto-WhitePaper2008,sasson2014zerocash,van2013cryptonote,Wood@Ethereum}. These smart contracts specify the terms of agreements between participants in the form of executable, tamper-proof code, removing the need for a trusted intermediary. This brings transparency, security, and decentralization to DeFi protocols, forming the backbone of many DeFi applications.

However, the processes through which we manage and amend these smart contracts, whether for fixing bugs, adapting to new requirements, or implementing upgrades, must also adhere to the principles of transparency and decentralization. This paper investigates whether the governance and voting mechanisms in practice, particularly in popular DeFi protocols such as Compound and Uniswap, align with these principles.

While there has been significant prior work exploring security vulnerabilities in smart contract implementations, particularly in DeFi protocols such as exchanges~\cite{adams2021uniswap,Daian@S&P20}, loans~\cite{Perez@FC21,Qin@FC21}, and auctions~\cite{NFTs}, few studies have examined vulnerabilities arising from the design and execution of the governance procedures that amend these smart contracts. In this study, we focus on analyzing the governance protocols of Compound and Uniswap over a 3-year period, from their inception until August 19, 2024 (block \#\num{20563000}), to identify potential risks and inefficiencies in the governance of DeFi smart contracts.

Governance protocols dictate how changes to smart contracts are proposed, discussed, and approved. These protocols aim to distribute decision-making power among the stakeholders of the protocol, with voting power typically proportional to a participant's economic stake. In theory, governance protocols should offer decentralized control and transparency, with decisions being made by stakeholders through a voting process. We empirically assess whether these properties hold in practice by analyzing the governance mechanisms employed by Compound and Uniswap, two widely-used protocols in the DeFi space.

\subsection{Relevance and Context of Decentralized Governance}
The governance of DeFi protocols, based on transparency and decentralization, contrasts sharply with the opaque, centralized governance structures seen in traditional centralized exchanges, such as FTX~\cite{Alameda@CoinDesk,Alameda@Investopedia}. In the case of FTX, a small group of centralized decision-makers undermined trust in the platform, resulting in its collapse. This failure highlights the importance of having governance mechanisms that allow for transparent and inclusive decision-making, ensuring that all stakeholders have a voice in critical decisions. Similar issues have arisen in traditional financial systems, such as the collapse of Silicon Valley Bank~\cite{SVB@WallStreet} and the de-pegging of the Luna token~\cite{Luna@CoinDesk,Luna@Forbes}, further underscoring the need for transparent decision-making in decentralized systems.

Although governance protocols offer a mechanism for decentralized decision-making, the mere presence of such systems does not guarantee that all stakeholder interests are adequately considered. This paper aims to answer three key questions by analyzing the governance practices of Compound and Uniswap: (i) \stress{Is voting power ``fairly'' distributed across stakeholders?}; (ii) \stress{Does the cost of voting serve as a deterrent to voter participation or act as an unfair ``polling tax''?}; and (iii) \stress{Do stakeholders collude or form coalitions to influence the outcomes of proposals?}

\subsection{Broader Implications of Governance}
Unlike centralized organizations where decisions are made by a small group of executives or a CEO, decentralized systems like DeFi protocols must align the incentives of individual stakeholders with the overall health and stability of the system. These challenges are not unique to blockchain; similar centralization problems have been observed in decentralized social media platforms like Mastodon~\cite{Mastodon} and Bluesky~\cite{Bluesky}, which face issues related to the cost of hosting, popularity of content, and centralization~\cite{raman2019challenges}. These challenges are analogous to issues observed in blockchain systems~\cite{Messias@IMC2021,Weintraub@IMC2022}.

The insights from this work, hence, extend beyond blockchain governance and offer valuable perspectives on the broader challenges faced by decentralized systems. By understanding how governance protocols operate within DeFi systems, we gain a deeper understanding of the complexities involved in building and maintaining truly decentralized applications.

\subsection{Key Contributions}

This paper makes key contributions to the understanding of governance in DeFi protocols, particularly, we show that:

\point{}\parai{Voting outcomes do not necessary capture the preferences of the broader community}
We analyze voting power distribution and find that a small group of top voters holds significant influence over the governance outcomes, with only a handful of voters able to determine the result of proposals. Specifically, 10 voters control \num{50.53}\% and \num{35.73}\% of the voting power and proposals only required an average of \num{3.18} and \num{4.7} voters to obtain at least \num{50}\% of the votes for Compound and Uniswap, respectively. This centralization of power is a critical issue, particularly in Compound where voting coalitions among top voters exacerbate the concern.

\point{}\parai{Voting is unfairly expensive for ``small'' stakeholders}
We show that voting costs in on-chain governance protocols are disproportionately high for smaller stakeholders. In Compound, voting costs ranged from \$\num{0.03} to \$\num{294.02}, with an average of \$\num{6.82}\footnote{All costs are in US dollars, taking into account the exchange rate at the time of casting the vote.}, while for Uniswap, the costs ranged from \$\num{0.17} to \$\num{126.86}, averaging \$\num{2.42}. When normalized by tokens held, the cost per vote unit is significantly higher for Compound (\$598.97) than for Uniswap (\$102.46), making it harder for smaller token holders to participate in governance.

\point{}\parai{Some voters cast their votes with significant delay}
We observe significant voting gaps, the time between the start of the voting period and when a vote is cast, with voters taking an average of \num{1.43} days in Compound and \num{2.72} days in Uniswap. These delays indicate strategic behavior, where voters withhold their votes to assess proposal viability before participating. This behavior aligns with the concept of voting sniping~\cite{feichtinger2024sok,yaish2024strategic}, where late-stage voting may influence outcomes at critical moments. Such timing strategies introduce vulnerabilities in governance, potentially undermining fairness and impeding quorum formation, especially in contentious proposals.

\point{}\parai{Scientific relevance}
This paper, in its preliminary and unpublished form, is one of the first to analyze voting power concentration in DAOs. Additionally, many subsequent and published studies have cited our work to discuss voting concentration~\cite{austgen2023dao,andres2025dao,Ma@ACM-Transactions,Motepalli@ICBC,tan2023open}, voter timing strategies~\cite{yaish2024strategic}, and various attacks related to DAOs~\cite{feichtinger2024sok,yaish2024strategic}. Thus, we believe our paper has made significant scientific contributions that are already recognized and validated by the research community.

\point{}\parai{Reproducibility and Data Availability}
We commit to promoting reproducible research by making all our scripts and data sets available through a public repository~\cite{Messias-Dataset-Code-2025}. This ensures that our findings can be verified by other researchers and supports continuous monitoring of governance protocols.

\section{Governance Mechanisms in Practice}
\label{sec:background}

The governance protocols that oversee the updating and modification of smart contracts are key to ensuring decentralized control and transparency. These protocols are designed to allow stakeholders to propose, vote on, and execute changes without relying on a central authority. In decentralized governance, voting power is often distributed among participants based on their economic stake in the protocol, with each participant's influence typically proportional to the amount of tokens they hold. This system is meant to ensure that all participants have an equal say in the decision-making process, thereby maintaining the decentralized nature of the protocol.

Our work aims at empirically analyzing two widely used governance protocols: Compound~\cite{leshner2019compound} and Uniswap~\cite{adams2021uniswap}. Both protocols have been instrumental in shaping the DeFi landscape and provide valuable case studies for understanding the real-world challenges of decentralized governance. By focusing on these protocols, we explore how governance decisions are made, how voting power is distributed, and the impact of voting costs and token delegation on the decentralization of governance. This delegation is a form of \stress{liquid democracy}, where voters can participate in decision-making either directly by voting or indirectly by delegating their voting rights to others~\cite{behrens2017origins,blum2016liquid,carroll1884principles}. The findings of this study contribute to the broader understanding of decentralized governance mechanisms and their implications for the future of blockchain-based applications.

\paraib{Compound}
It is a decentralized lending protocol that allows users to lend and borrow tokens or assets via smart contracts.
Lenders earn interest (\stress{yield}) by supplying liquidity to the protocol, while borrowers obtain tokens from the protocol and pay interest on the borrowed tokens.
Compound protocol has 2 versions of its governance contract: \stress{Alpha} and \stress{Bravo}.
\stress{Compound Governor Alpha}, the first version of the governance contract, was deployed on Mar. 4\tsup{th}, 2020 (block \#\num{9601459}) and was active until Mar. 28\tsup{th}, 2021 (block \#\num{12126254}).
The upgrade, \stress{Compound Governor Bravo}, was deployed on Mar. 9\tsup{th}, 2021 (block \#\num{12006099}) and has been active since Apr. 14\tsup{th}, 2021 (block \#\num{12235671}).
Bravo introduced several improvements such as smart-contract upgradability (through proxies), a new option for voters to abstain from voting, and the ability for voters to state the reasons behind their voting choices through text comments attached to on-chain votes.
The Bravo contract was proposed in proposal \#\num{42}, and it received \num{1438679.86} votes from \num{59} voters---all but one vote was in favor of its implementation.

\paraib{Uniswap}
It is a decentralized exchange (DEX) protocol that enables users to trade one token for another.
In addition, users can add liquidity to Uniswap and earn interest based on the amount they provide.
The Uniswap token (UNI) is an ERC-20 token that allows its holders to transfer UNI between addresses.
Similarly to Compound, with Uniswap, holders can also participate in the Uniswap Governance Protocol by proposing changes and voting on proposals.
Uniswap Governance is derived from Compound Governance Alpha and Bravo deployed on May 31\tsup{st}, 2021 (\#\num{12543659}) and August 20\tsup{th}, 2021 (\#\num{13059157}), respectively. Therefore, it shares similar features but has some particularities, for example, the voting period in Uniswap is longer than in Compound~\cite{Uniswap-Lifecycle} and it allows for whitelisting members to allow them to submit proposals even with 0 tokens delegated to their addresses.

\subsection{Voting Modalities}

In decentralized governance protocols, voting modalities define the rules and mechanisms through which stakeholders participate in decision-making. These modalities can significantly impact the decentralization, efficiency, and fairness of governance.

\paraib{On-Chain Voting}
On-chain voting refers to governance decisions that are made directly on the blockchain. In this modality, participants cast their votes by issuing transactions on the blockchain, which are recorded and verified as part of the public ledger. This process ensures transparency and immutability, as all votes are publicly visible and cannot be altered once committed. On-chain voting also often requires participants to pay transaction fees to submit their votes, which can vary based on network congestion. Well-known DeFi protocols like Compound and Uniswap use on-chain voting to allow token holders to vote on proposals related to protocol upgrades and changes. In these systems, voting power is typically proportional to the number of tokens a participant holds, with some protocols also allowing token delegation for those who do not wish to vote directly.

\paraib{Off-Chain Voting}
Off-chain voting, on the other hand, occurs outside the blockchain, typically on third-party platforms such as Snapshot~\cite{Snapshot} and Tally~\cite{Tally}. In this modality, votes are not recorded on the blockchain but are instead collected and aggregated by a separate service. While off-chain voting eliminates the need for transaction fees, it may introduce certain vulnerabilities, such as the potential for centralization or manipulation by the platform administrators. After the voting period, the results are typically executed on-chain through multisig transactions or other mechanisms. Off-chain voting systems are often used for less critical decisions or when the cost of on-chain voting is prohibitively high. For example, protocols like Balancer~\cite{Governance@Balancer} and Convex Finance~\cite{Governance@ConvexFinance} use off-chain voting to handle governance decisions. Uniswap, on the other hand, employs off-chain voting as a preliminary evaluation of proposals before they are officially proposed on-chain.

\paraib{Delegated Voting}
In many governance protocols, participants can delegate their voting power to other participants. This modality enables a form of liquid democracy, where token holders who do not want to vote directly can empower others to cast votes on their behalf~\cite{behrens2017origins,blum2016liquid,carroll1884principles}. Delegation can either be permanent or temporary, and it is often used to concentrate voting power among a smaller number of trusted individuals or entities. For instance, Compound and Uniswap allow token holders to delegate their votes to other users, enabling a more flexible and efficient governance process.

\section{Methodology}
\label{sec:methodology}

\begin{figure}[t]
\centering
\includegraphics[width=0.7\textwidth]{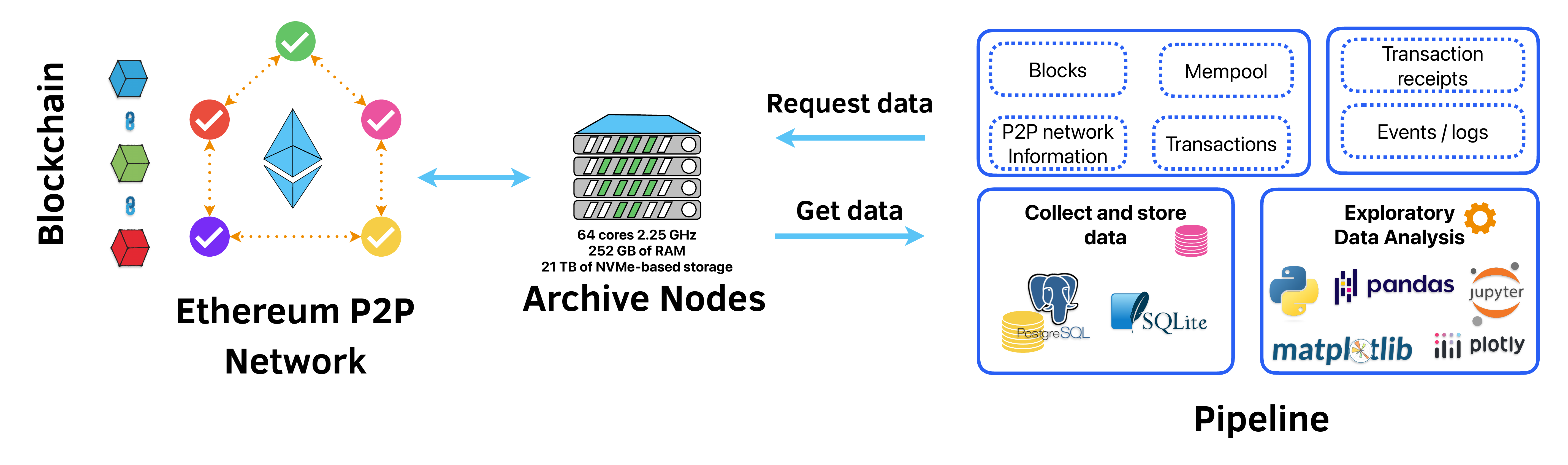}
\figcap{Overview of the data collection methodology.}\label{fig:data-pipeline}
\end{figure}

In this section, we describe the methodology used to empirically analyze the decentralization, fairness, and governance dynamics of the Compound and Uniswap protocols.

\subsection{Data Collection Strategy}
\label{sec:data-collection}
We gather data directly from the Ethereum blockchain using an Ethereum archive node. Specifically, we collect events related to token transfers, delegation changes, proposal creation, and vote casting for Compound and Uniswap. The data spans from the inception of each protocol to August 19, 2024, covering Compound from March 4, 2020 (block \#\num{9601459}) and Uniswap from August 20, 2021 (block \#\num{13059157}). We present our data set in Table~\ref{tab:events_tokens_governance}.

The data collection process involves setting up an Ethereum archive node on a server with 64 cores, 252 GB of RAM, and 21 TB of storage, which synchronizes with the Ethereum blockchain. Using the Web3.py Python library, we query and retrieve relevant governance-related events for both protocols. This dataset includes token transactions, governance proposals, voting activity, and delegation details~\cite{Messias-Dataset-Code-2025}. We illustrate our data-analysis pipeline in Figure~\ref{fig:data-pipeline}.

Furthermore, since blockchain wallet addresses are typically anonymous, we infer the ownership of addresses to provide a more accurate analysis of the governance process. To do this, we leverage publicly available data sources where users may choose to disclose their identities, such as Etherscan~\cite{Etherscan@ETH-explorer}, Sybil-List~\cite{Addresses@Sybil}, and Tally~\cite{Tally}. By matching Ethereum wallet addresses from these sources, we can identify the real-world entities controlling certain addresses, allowing us to cluster addresses controlled by the same entity. This ownership inference helps improve the accuracy of our analysis by associating specific voting and governance actions with identifiable entities, thereby enhancing the understanding of governance dynamics in Compound and Uniswap.

For Compound, we successfully matched the owners of 22 out of 45 unique addresses (\num{48.89}\%) associated with proposal creation and 339 out of \num{4538} proposal voters (\num{7.47}\%). For Uniswap, we matched the owners of 15 out of 31 unique addresses (\num{48.39}\%) linked to proposal creation and 458 out of \num{20695} proposal voters (\num{2.21}\%).

Additionally, we matched the ownership of 101 addresses (\num{47.87}\%) of the top 10 most influential voters in Compound and 45 addresses (\num{67.16}\%) of the top voters in Uniswap, as determined by the number of delegated tokens they controlled when casting their vote. This reveals that nearly half of the most influential voters in both protocols chose to disclose their identities.

Finally, we group the identified addresses belonging to the same entity to conduct a more accurate analysis of voting patterns and governance dynamics.

\begin{table*}[tb]
\centering
\small
\caption{Summary of events related to the Compound (COMP), Uniswap (UNI) tokens, and their Governor contracts that we gathered from the Ethereum blockchain from their inception to Aug. 19, 2024 (block \#\num{20563000}).}
\label{tab:events_tokens_governance}
\resizebox{\textwidth}{!}{%
\begin{tabular}{rrrrp{6cm}}
\toprule
\multirow{2}{*}{\thead{Contract type}} &\thead{Event name} & \multicolumn{2}{c}{\thead{\# of events}} & \thead{Description} \\
        &   & \thead{Compound}        & \thead{Uniswap}        &             \\
\midrule
\multirow{4}{*}{\thead{Token}} & \stress{Approval} & \num{256032} & \num{798107}   & \small{Standard ERC-20 approval event.} \\
 & \stress{DelegateChanged} & \num{15244} & \num{53977}   & \small{Emitted when an account changes its delegate.} \\
 & \stress{DelegateVotesChanged} & \num{83798} & \num{163347}    &\small{Emitted when a delegate account's vote balance changes.} \\
 & \stress{Transfer} & \num{2224966} &  \num{4765566}  & \small{Emitted when users/holders transfer their tokens to another address.} \\
\hline
\multirow{5}{*}{\thead{Governance}} & \stress{ProposalCanceled} & \num{30} &  \num{14}   &  \small{Emitted when a proposal is canceled.} \\
 & \stress{ProposalCreated} & \num{307} & \num{67}    &  \small{Emitted when a new proposal is created.} \\
 & \stress{ProposalExecuted} & \num{238} &  \num{43}   &  \small{Emitted when a proposal is executed in the TimeLock.} \\
 & \stress{ProposalQueued} & \num{246} &  \num{43}   &  \small{Emitted when a proposal is added to the queue in the TimeLock.} \\
 & \stress{VoteCast} & \num{14841} &  \num{51580}    &  \small{Emitted when a vote is cast on a proposal: \num{0} for against, \num{1} for in-favor, and \num{2} for abstain.} \\
\bottomrule
\end{tabular}}
\end{table*}

\subsection{Experimental Design}
\label{sec:experimental-setup}
We focus on two widely used decentralized governance protocols: Compound and Uniswap. These protocols are chosen because of their significance in decentralized finance (DeFi) and their similar design principles, including token-based governance and on-chain voting. Our goal is to analyze the distribution of voting power among protocol participants, voting participation patterns, and the impact of token delegation. We also investigate the cost of voting, which includes transaction fees associated with on-chain voting, and explore the temporal dynamics of governance proposal outcomes.

To achieve this, we examine governance events in both protocols, such as token transfers, delegate changes, proposal creation, voting behavior, and proposal execution. We place particular emphasis on analyzing the concentration of voting power and its implications for decentralization. By focusing on these key aspects, we aim to understand how decentralized governance works in practice and whether it aligns with the theoretical promises of decentralization.

\subsection{Analysis Methods}
\label{sec:analysis-techniques}

In this section, we outline our analysis techniques used to extract meaningful insights from the data collected in our study.

\paraib{Token Distribution and Delegation Analysis}
We analyze the distribution of governance tokens in both Compound and Uniswap to assess the centralization of voting power. We investigate how tokens are distributed among the top token holders and evaluate the effects of token delegation on governance. Delegation allows users to transfer their voting power to others, which can result in concentrated decision-making power. By analyzing token delegation data, we assess how this mechanism affects the decentralization and fairness of governance processes. Specifically, we explore the extent to which delegation leads to centralization and whether the power to propose or vote on changes is concentrated in a few hands.

\paraib{Voting Cost Analysis}
We analyze the cost of voting in both protocols in US dollars, using the ETH-USD Yahoo Finance data feed~\cite{eth-usd@yahoo} exchange rate, which varies depending on Ethereum network congestion and transaction fees. Specifically, we focus on the transaction fees associated with casting votes and calculate the average cost per vote for both protocols. To better understand the fairness of voting, we normalize the voting costs by the number of tokens held by each voter in separate analyses. This provides insight into the per-unit cost of casting a vote, especially for smaller stakeholders who may face disproportionately high fees. By assessing the financial barriers to voting, we can evaluate whether these costs affect the accessibility and fairness of governance participation.

\paraib{Temporal Dynamics of Proposal}
We investigate how proposals evolve throughout their lifecycle by analyzing the frequency, timing, and outcomes of governance proposals in Compound and Uniswap. We begin by identifying the set of proposers, quantifying how often proposals are submitted, and measuring the intervals between proposals. We then assess the lifecycle of each proposal, including the creation time, voting period, and execution timeline, to capture differences in protocol-level governance design (e.g., proposal delays, voting windows, timelocks). Additionally, we track the final outcomes of all proposals (executed, defeated, or canceled) and identify when and why cancellations occur (e.g., before voting, during the timelock, or after approval). This methodology allows us to characterize the governance cadence of each protocol and detect anomalies or bottlenecks that may reflect coordination challenges or power dynamics.

\paraib{Cosine Similarity for Voting Patterns}
To investigate the formation of voting coalitions, we employ cosine similarity~\cite{Cosine@ScikitLearn,xia2015learning} to measure how similarly voters cast their votes on governance proposals. The cosine similarity value ranges from \num{-1} to \num{1}, with a value of \num{1} indicating a high degree of similarity. It quantifies the degree to which two voting patterns align, which helps us identify potential voting groups or coalitions.
This metric enables us to detect coordinated voting behaviors, which may indicate centralization risks within the governance process. By comparing voting patterns, we can uncover whether a small number of entities exert undue influence over governance outcomes, potentially undermining decentralization.

\section{Power Dynamics in Blockchain Governance}\label{sec:compound}

In this section, we analyze two popular governance protocols: Compound and Uniswap. Since many other governance protocols are based on the Compound Governor, our work can serve as a guide for other protocols that forked from Compound's architecture.

\subsection{Voting Power Distribution Analysis}

The distribution of voting power plays a significant role in the decentralization of governance. Voting power is typically proportional to the amount of governance tokens a user holds, with each token representing one vote.

\begin{figure*}[tb]
    \centering
    \begin{subfigure}{\twocolgrid}
        \centering
        \includegraphics[width=\twocolgrid]{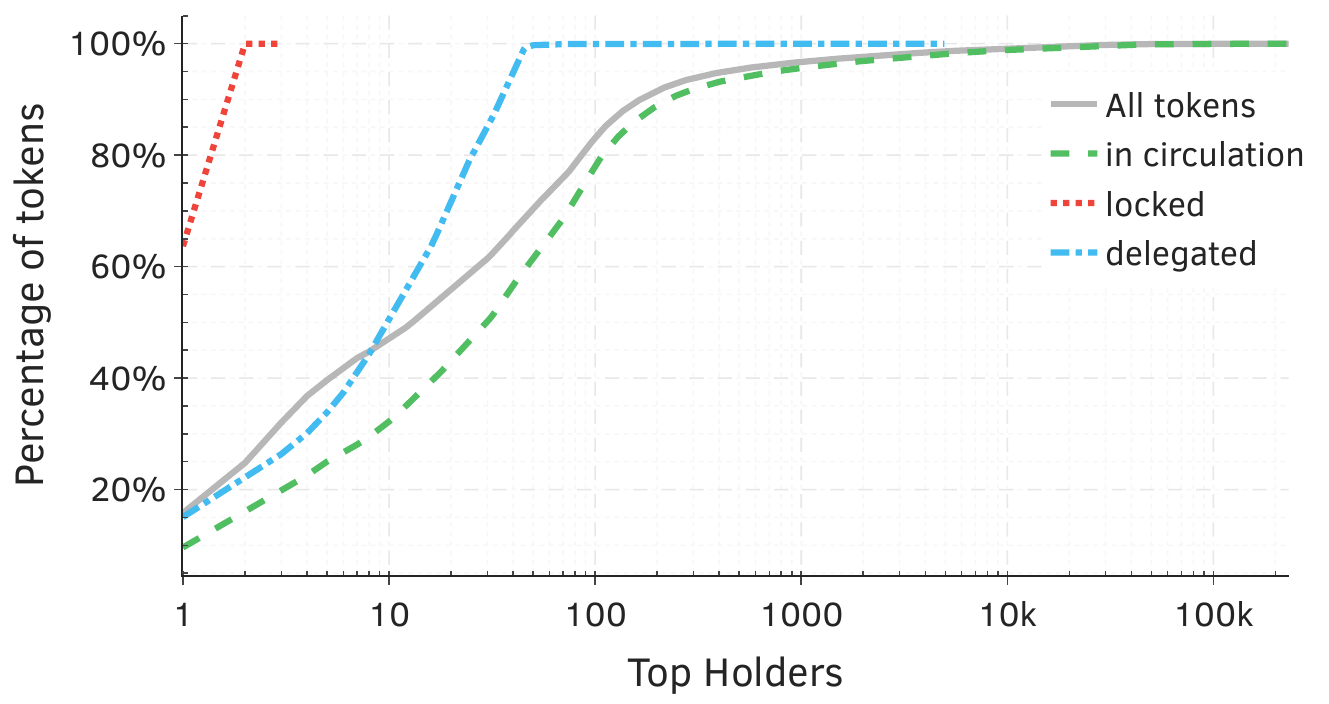}
        \caption{Compound}
        \label{fig:new_compound-hold-tokens-comp}
    \end{subfigure}
    \begin{subfigure}{\twocolgrid}
        \centering
        \includegraphics[width=\twocolgrid]{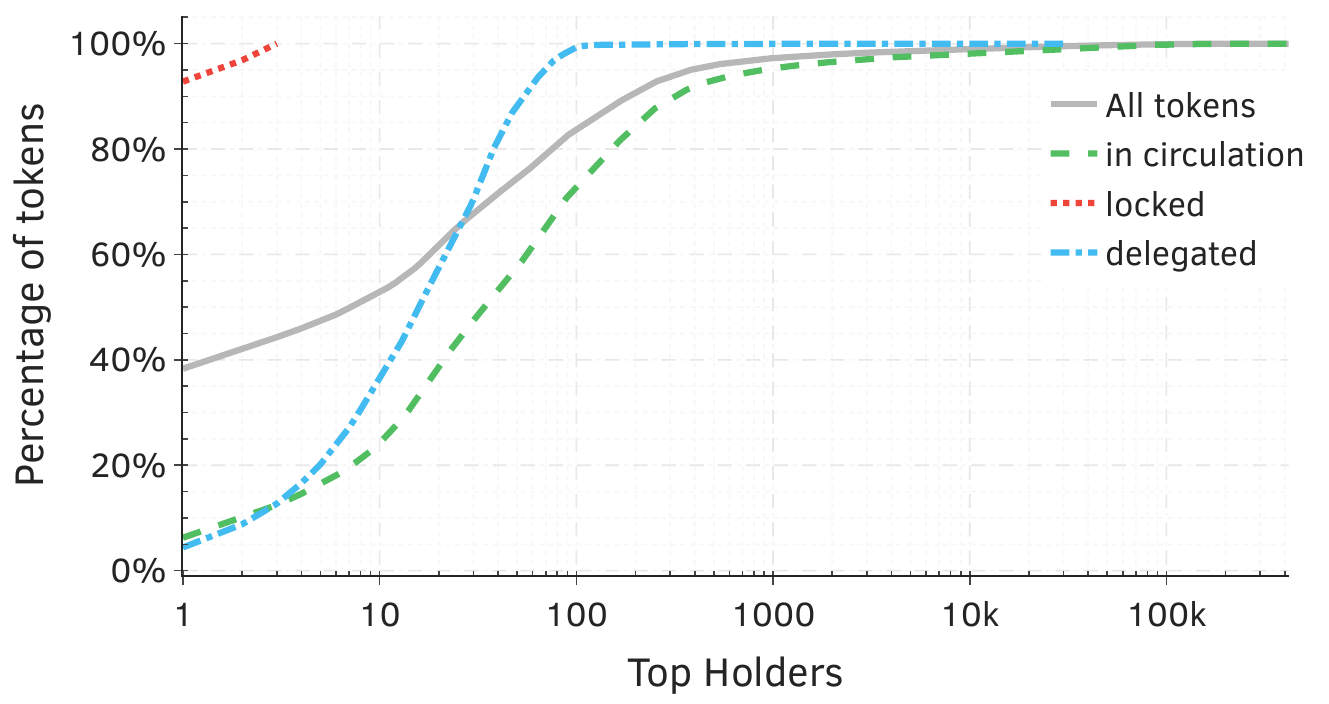}
        \caption{Uniswap}
        \label{fig:new_uniswap-hold-tokens-uni}
    \end{subfigure}\hfill
    \caption{
    CDF of the fraction of COMP (UNI) tokens held per account. The 10 million (1 billion) tokens available are shared among 231,396 (417,551) accounts, in \grey{grey}. The dashed \green{green} line shows the distribution of the fraction of 7.5 million (587.11 million) accounting for 75.20\% (58.71\%) of the tokens in circulation held by 231,393 (417,548) accounts. The remaining tokens are locked and held by 3 accounts (in dotted \red{red}). The dash-dotted \blue{blue} line shows the delegated tokens' distribution where 10 out of 4944 (32013) accounts have 50.53\% (36.24\%) of all delegated tokens available.
    }
    \label{fig:new_compound-uniswap-hold-tokens}
\end{figure*}

\subsubsection{Token Distribution}

The initial token distributions for both protocols were designed to encourage active participation. In Compound, 42.15\% of the total supply (10 million COMP tokens) was allocated to liquidity mining, 23.95\% to shareholders, 22.46\% to the founders and team, 7.73\% to the community, and 3.71\% to future team members~\cite{Compound@CoinGecko}. However, the public release of COMP tokens only began after proposal \#7 was executed on June 15, 2020. This proposal enabled the continued distribution of tokens to users over time. As of our analysis, 10 million COMP tokens have been distributed across 231,396 accounts. The largest holder, \textit{Compound Comptroller}, holds 15.79\% (\num{1578590.06} tokens), followed by Compound USDCv3 Token (9\% or \num{899995.09} tokens) and Binance (7.23\% or \num{722734.90} tokens).

However, only 7.5 million of these tokens are actively in circulation, and we show their distribution in Figure~\ref{fig:new_compound-hold-tokens-comp}. The tokens in circulation are those available for trading or exchange, excluding \textit{locked} tokens from the Compound Reservoir, Comptroller, and Timelock contracts, which are not in circulation~\cite{Governance@Compound, kybx86@Compound}. These locked tokens require governance approval for release, though some are periodically distributed as incentives for protocol use, such as lending or borrowing.

Similarly, Uniswap’s initial token distribution consisted of 1 billion UNI tokens. 60\% of these were allocated to the community, 21.27\% to Uniswap's team, 18.04\% to investors, and 0.69\% to advisors~\cite{Uniswap@CoinGecko}. By August 2024, the 1 billion UNI tokens were distributed among \num{417551} accounts. The largest holder, \textit{Uniswap Timelock}, controls 38.31\% of the total supply, followed by Binance, with 3.67\% of the total supply.

Moreover, at the time of our analysis, the distribution of tokens in both Compound and Uniswap reveals a significant concentration among a small number of top holders. In Compound, the top 15 holders control 38.09\% of all tokens in circulation (Figure~\ref{fig:new_compound-hold-tokens-comp}), while in Uniswap, the top 15 control 31.87\% (Figure~\ref{fig:new_uniswap-hold-tokens-uni}). This concentration of tokens is further amplified by Binance, a centralized exchange, which is the largest token holder in both protocols, controlling 9.61\% of Compound’s and 6.25\% of Uniswap's total circulation supply. Although Binance theoretically could delegate these tokens, it has stated that it will not use them to vote on behalf of its users~\cite{Binance-denies@coindesk}.

\begin{mdframed}[style=Takeaways]
\textbf{Takeaway:} Both Compound and Uniswap exhibit a significant concentration of token holdings among a small group of top holders, with centralized entities like Binance holding a substantial share.
\end{mdframed}

\subsubsection{Token Delegation}

Delegation is a prerequisite for voting in most governance protocols, including Compound and Uniswap, allowing participants to transfer their voting rights to others. This ability enables users to delegate their voting power to individuals who share their interests, allowing smaller participants to pool their votes together and have a more significant impact. However, users can only delegate \textit{all} their tokens, not a fraction, of them. The protocol enforces this limitation at the wallet address level. Users can own multiple wallet addresses and divide their tokens among them, thereby enabling delegation of a subset of their tokens to others~\cite{Governance@a16z, fritsch@2022votingpower}.

To determine whether delegated tokens are held by a few voters, we group together all inferred addresses (as discussed in~\S\ref{sec:data-collection}) that belong to the same entity and then calculate the total number of delegated tokens (i.e., voting power) held by each group. From this analysis, we observe that delegated tokens are highly concentrated among a few voters (refer to the dash-dotted blue line in Figure~\ref{fig:new_compound-uniswap-hold-tokens}).

In Compound, out of \num{4944} delegatee (accounts with voting rights), the top 50 (1.01\%) hold 99.71\% of all delegated tokens. Similarly, in Uniswap, the top 50 (0.16\%) out of \num{32013} delegatee accounts hold 88.28\% of all UNI delegated tokens. On August 19, 2024, a16z held the most COMP delegated tokens, with 15.06\% (\num{361008} tokens), followed by the address 0x7e95$\cdots$1318 with 7.09\% (\num{170000} tokens), and Geoffrey Hayes with 4.21\% (\num{101008} tokens). These three addresses together held 26.36\% (\num{632016} tokens) of the total \num{2397812.83} delegated tokens in our analysis. Notably, only 33.51\% and 34.89\% of the tokens in circulation are delegated in Compound and Uniswap, respectively.

\begin{mdframed}[style=Takeaways]
\textbf{Takeaway:} Token delegation in both Compound and Uniswap is highly concentrated among a small number of large entities, which further centralizes governance power and diminishes the decentralization of decision-making processes.
\end{mdframed}

\subsubsection{Voting Power Concentration}

\begin{figure*}[tb]
\centering
\begin{subfigure}{\twocolgrid}
  \centering
  \includegraphics[width=\twocolgrid]{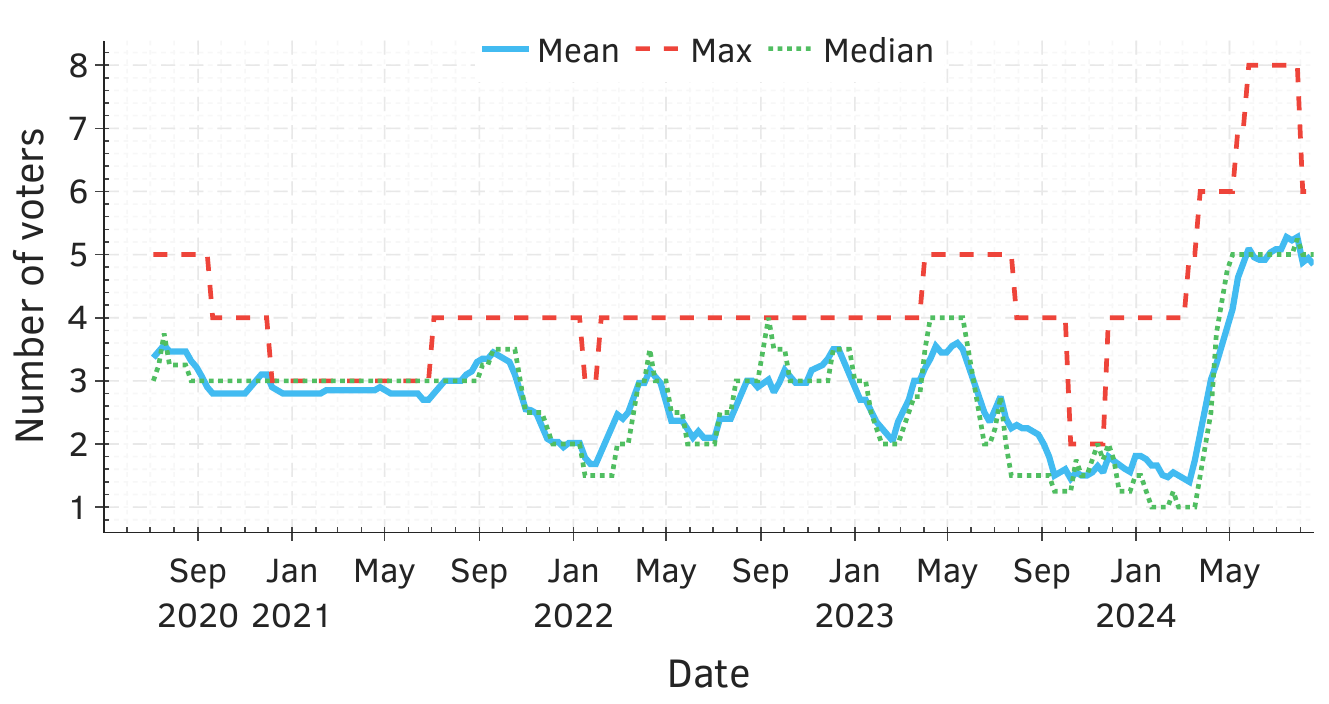}
  \caption{Compound}
\label{fig:voting_power_to_reach_50_perc_of_votes_compound}
\end{subfigure}
\begin{subfigure}{\twocolgrid}
  \centering
  \includegraphics[width=\twocolgrid]{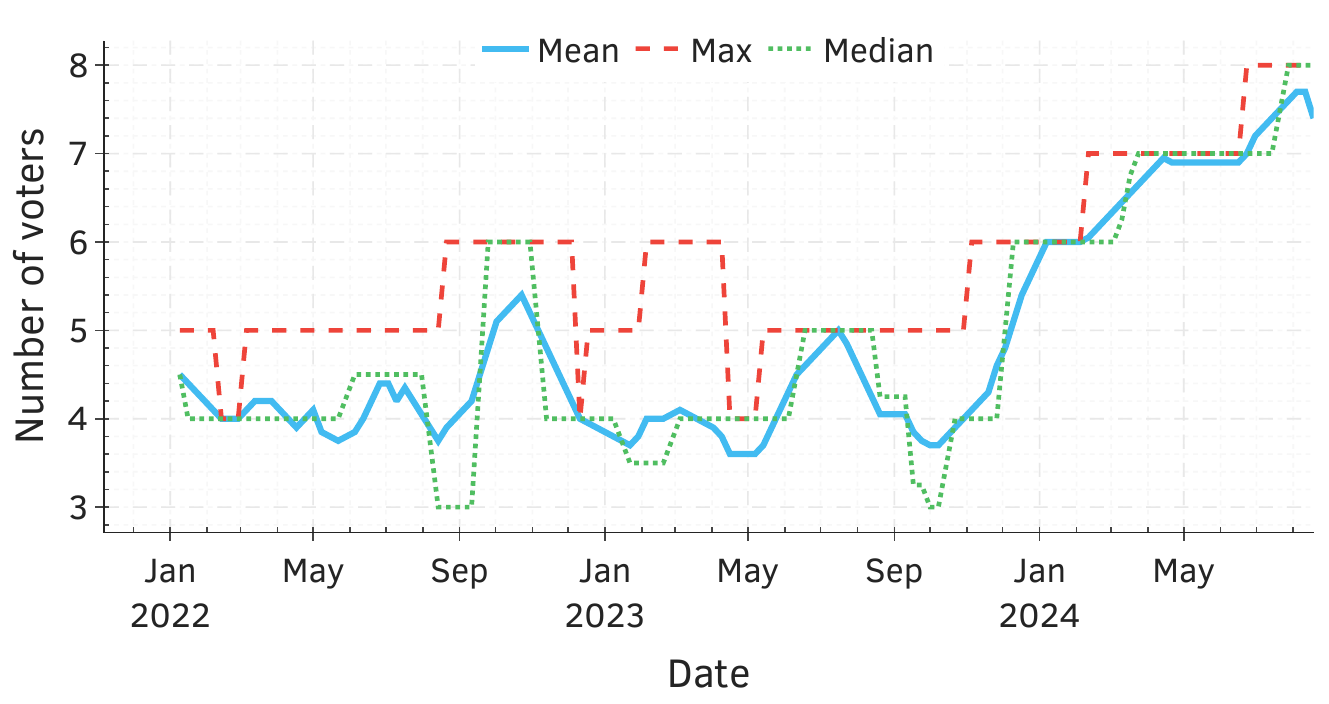}
  \caption{Uniswap}
\label{fig:voting_power_to_reach_50_perc_of_votes_uniswap}
\end{subfigure}
\caption{Minimum number of voters required to reach at least 50\% of the total votes in each proposal: (a) Compound proposals required an average of 3.18 voters. The median was 3 voters, with a range from 1 to 8. (b) Uniswap proposals required an average of 4.7 voters. The median was 5 voters, with a range from 1 to 8. This indicates concentrated voting power in these protocols.}
\label{fig:voting_power_to_reach_50_perc_of_votes}
\end{figure*}

We analyzed the number of voters required to reach at least 50\% of the total votes cast for each proposal, often referred to as the \stress{Nakamoto Coefficient} when applied to decentralized autonomous organizations (DAOs)~\cite{wu2020coefficient}.

For the \num{307} Compound proposals, we excluded \num{15} proposals canceled before the voting period began, leaving \num{292} (\num{95.11}\%) proposals for analysis. Similarly, for Uniswap, we considered \num{50} (\num{74.63}\%) proposals in our analysis.
On average, Compound proposals required \num{3.18} voters (std. of \num{1.54}) to reach at least 50\% of the votes, with a median of \num{3} voters. The number of voters ranged from \num{1} to \num{8}. We provide the details in Figure~\ref{fig:voting_power_to_reach_50_perc_of_votes_compound}.
In Uniswap, on average, proposals required \num{4.7} voters (std. of \num{1.59}) to reach at least 50\% of the total votes, with a median of \num{5} voters. The minimum number of voters was \num{1}, while the maximum was \num{8}. See Figure~\ref{fig:voting_power_to_reach_50_perc_of_votes_uniswap} for details.

These results suggest that voting power in both protocols is highly concentrated among a small number of voters who hold a significant proportion of the voting tokens.

\begin{mdframed}[style=Takeaways]
\takeaways{Takeaway:}
Our analysis shows that voting power in both Compound and Uniswap is highly concentrated, with only a small number of voters required to reach at least 50\% of the total votes, indicating significant control by a few token holders.
\end{mdframed}

\subsubsection{Coalition Formation in Voting}

\begin{figure*}[t]
    \centering
    \begin{subfigure}{\twocolgrid}
        \centering
        \includegraphics[width=\twocolgrid]{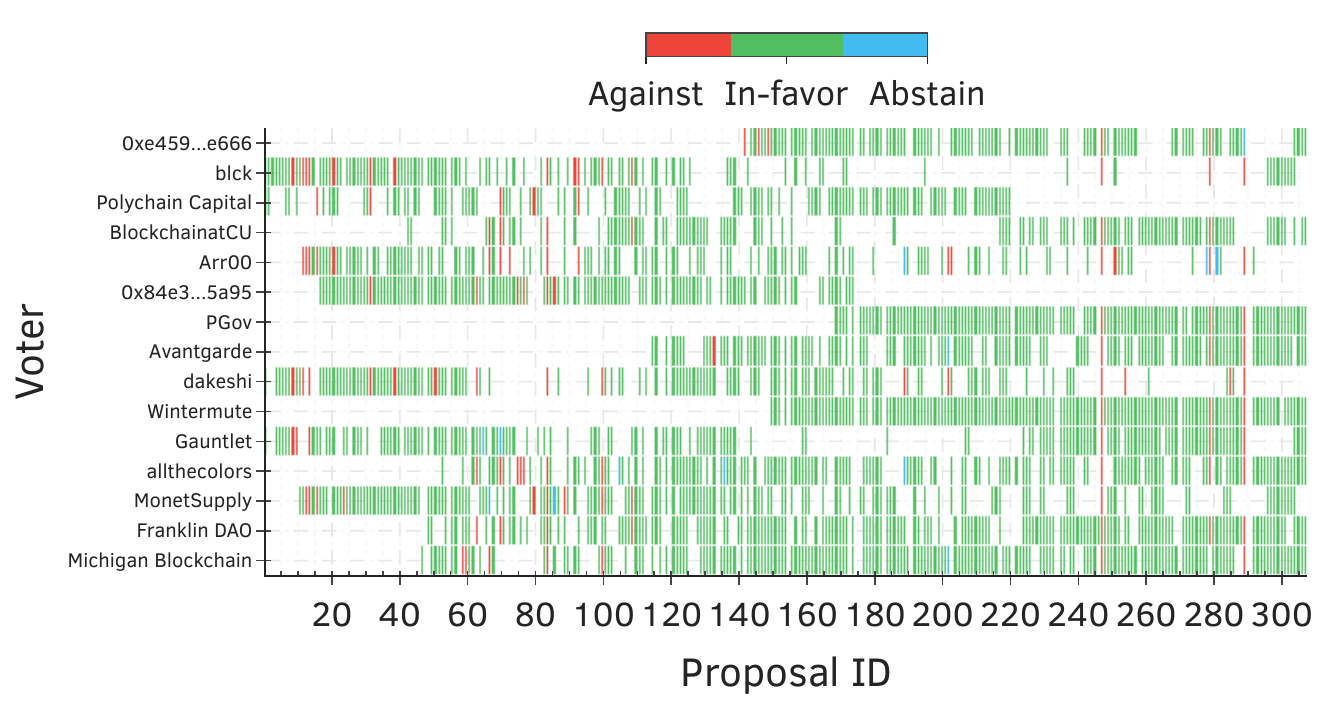}
        \caption{Compound}
        \label{fig:new_comp-votes-top-15-voters}
    \end{subfigure}
    \begin{subfigure}{\twocolgrid}
        \centering
        \includegraphics[width=\twocolgrid]{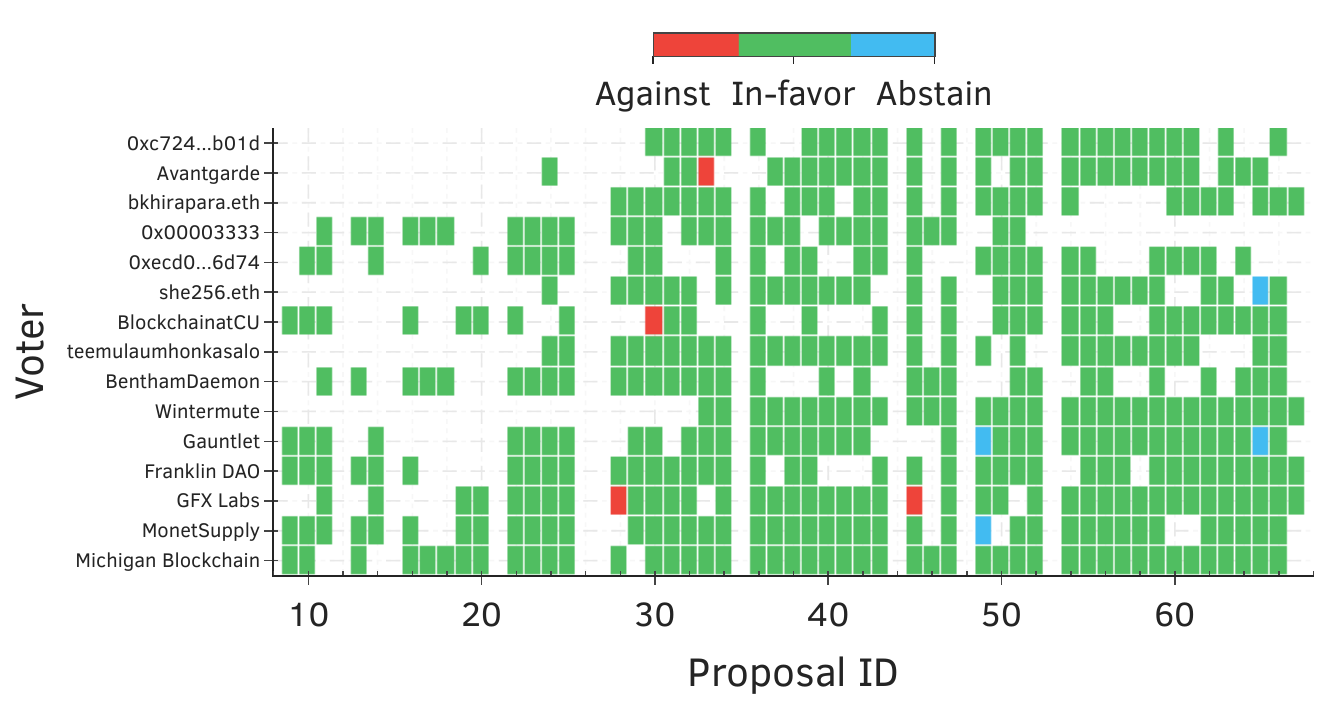}
        \caption{Uniswap}
        \label{fig:new_uni-votes-top-15-voters}
    \end{subfigure}
    \caption{
    Votes cast by the top-15 most frequently Compound and Uniswap voters. In-favor votes are in \green{green}, against in \red{red}, and abstain in \blue{blue} color.
    }
    \label{fig:new-votes-top-15-voters}
\end{figure*}

\begin{figure*}[t]
    \centering
    \begin{subfigure}{\twocolgrid}
        \centering
        \includegraphics[width=\twocolgrid]{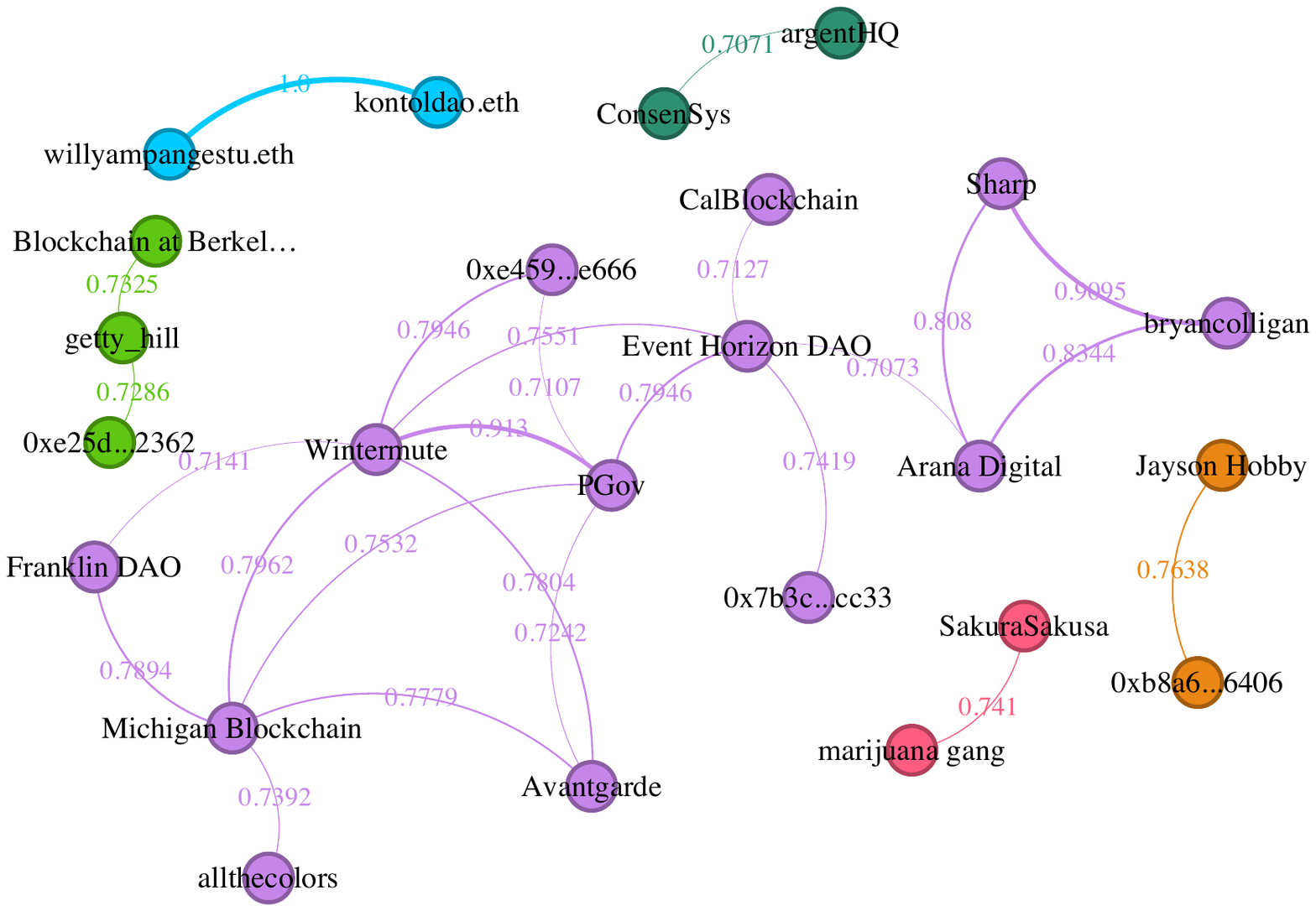}
        \caption{Compound}
        \label{fig:similarity-network-compound}
    \end{subfigure}
    \begin{subfigure}{\twocolgrid}
        \centering
        \includegraphics[width=\twocolgrid]{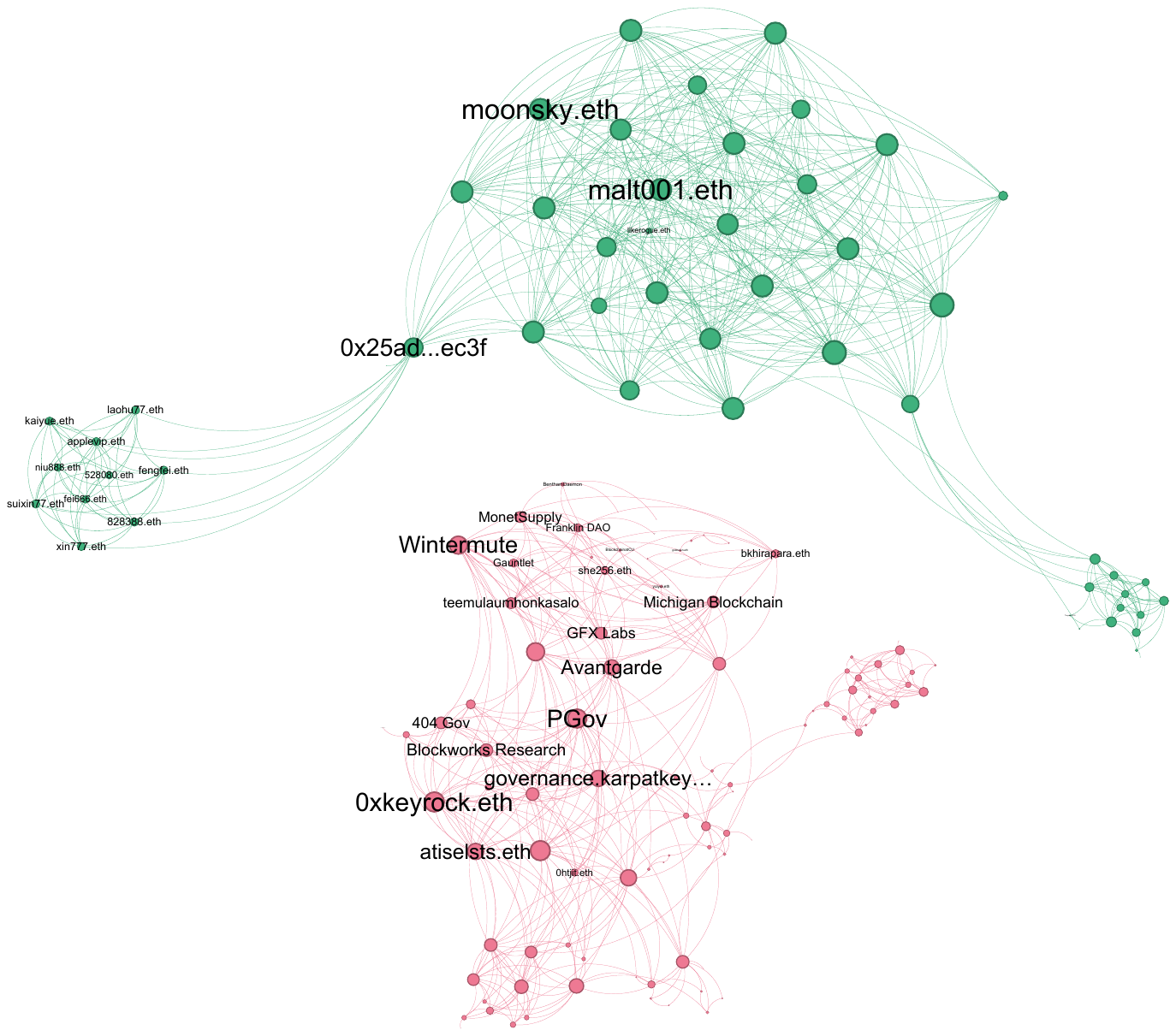}
        \caption{Uniswap}
        \label{fig:similarity-network-uniswap}
    \end{subfigure}
    \caption{
    Voter similarity graphs (cosine similarity $\geq$ 0.70), where nodes represent voters and edges indicate high similarity in their voting behavior. Node colors denote connected components. (a) Compound: 6 components, 24 nodes, 25 edges; one large cluster of 13 voters. Notably, the purple component contains 13 voters exhibiting a high degree of mutual similarity. (b) Uniswap: 2 large components with 150 nodes and 806 edges, showing stronger voter clustering.
    }
    \label{fig:similarity-network}
\end{figure*}

The concentration of voting power is influenced not only by the distribution of tokens but also by the formation of voter coalitions. To explore coalition formation, we examine the behavior of voters who may cast their votes collectively rather than individually. Such behavior can compromise the security of the governance protocol, as voters might imitate the choices of their peers rather than express independent preferences. The transparency provided by the Ethereum blockchain, which allows public access to voter addresses along with their voting power and preferences, may facilitate this type of behavior during the election process.

This analysis is essential, as it helps in understanding the voting decision-making patterns. Figure~\ref{fig:new_comp-votes-top-15-voters} presents a heatmap of the top \num{15} Compound voters, based on the number of proposals they have voted on or participated in. A similar analysis for Uniswap is shown in Figure~\ref{fig:new_uni-votes-top-15-voters}. We see on both figures that voters tend to cast similar votes on proposals except for a few that votes differently.

We also analyzed vote alignment by comparing each voter's choice to the final outcome of the proposal. Specifically, we define an \stress{aligned vote} as one cast \stress{against} a proposal that was ultimately defeated, or \stress{in favor} of a proposal that was eventually executed. This analysis offers insights into how often voters were able to achieve their desired outcomes. We consider only voters who participated in at least 15 proposals.
In Compound, we found that, on average, a voter's decisions aligned with the proposal outcomes 88.59\% of the time (standard deviation: 0.05\%, median: 88.05\%), with individual alignment rates ranging from 76.20\% to 96.67\%. This suggests that voters were generally effective in casting votes aligned with their preferences.
In Uniswap, the average alignment was even higher, at 91.45\% (standard deviation: 0.05\%, median: 90.32\%), with alignment rates ranging from 86.7\% to 97.56\%. These findings indicate that, in both protocols, voters tended to cast votes that were useful in achieving their intended outcomes.

Additionally, to better understand voter alignment in DAO governance, we construct cosine similarity (refer~\S\ref{sec:analysis-techniques}) graphs based on historical voting behavior, using a threshold of 0.70 to indicate strong alignment. In Compound, per Figure~\ref{fig:similarity-network-compound}, the resulting network is fragmented, with 24 voters forming 6 disconnected components and only 25 high-similarity edges. The largest component contains 13 voters with closely aligned preferences, suggesting the presence of a small, cohesive voting bloc amid otherwise diverse or uncoordinated participants. In contrast, the Uniswap similarity graph reveals a substantially denser and more cohesive structure: 150 voters are organized into only two components, connected by 806 similarity edges (see Figure~\ref{fig:similarity-network-uniswap}). This indicates stronger alignment in voting behavior, possibly due to influential delegates or better coordination mechanisms. These differences suggest that while Compound exhibits fragmented voter behavior, Uniswap governance is characterized by more centralized or coordinated voting dynamics.

\begin{mdframed}[style=Takeaways]
\takeaways{Takeaway:} 
Our analysis reveals that similar voting patterns among top voters in Compound and Uniswap suggest the formation of voting blocs, which could threaten decentralization by concentrating decision-making power even further.
\end{mdframed}

\subsection{Temporal Dynamics of Proposals}

In this section, we discuss the creation of proposals, the frequency at which proposals are submitted, their lifecycle, and their outcome.

\subsubsection{Proposal Submission Process}

To submit a proposal for a governance change, an address must have a minimum number of tokens delegated to it. Both Compound and Uniswap require at least 0.25\% of the total token supply delegated to the proposing address. Specifically, this corresponds to \num{25000} COMP tokens for Compound and \num{2500000} UNI tokens for Uniswap.

In Compound, \num{45} distinct proposers initiated \num{307} proposals during the observation period. Of these proposers, \num{18} (\num{40}\%) created only one proposal, while \num{26.67}\% submitted at least \num{8} proposals. On average, each proposer submitted \num{6.82} proposals, with a standard deviation of \num{13.23} and a median of \num{2}. 
The highest number of proposals was submitted by \stress{Gauntlet}, which created \num{83} proposals, followed by \stress{0x7b3c$\cdots$cc33} with \num{32} proposals. Gauntlet, hired by Compound to manage protocol risk, frequently submits proposals for adjusting the protocol's parameters~\cite{Gauntlet@Compound}.

In contrast, Uniswap saw \num{31} distinct proposers who initiated \num{67} proposals. Of these proposers, \num{14} (\num{45.16}\%) created just one proposal, while \num{2} (\num{6.45}\%) submitted at least \num{8} proposals. On average, each proposer submitted \num{2.16} proposals, with a standard deviation of \num{2.1} and a median of \num{2}. The proposer who submitted the most proposals was \stress{GFX Labs}, with a total of \num{10} proposals, followed by \stress{Michigan Blockchain}, with \num{9} proposals.

\begin{mdframed}[style=Takeaways]
\takeaways{Takeaway:}
To propose governance changes, both Compound and Uniswap require at least 0.25\% of the total token supply delegated to the proposing address. In Compound, proposers are more active, with a higher average number of proposals per proposer (6.82), while in Uniswap, fewer proposals are initiated on average (2.16).
\end{mdframed}

\subsubsection{Proposal Frequency Trends}

\begin{figure*}[t]
    \centering
    \begin{subfigure}{\twocolgrid}
        \centering
        \includegraphics[width=\twocolgrid]{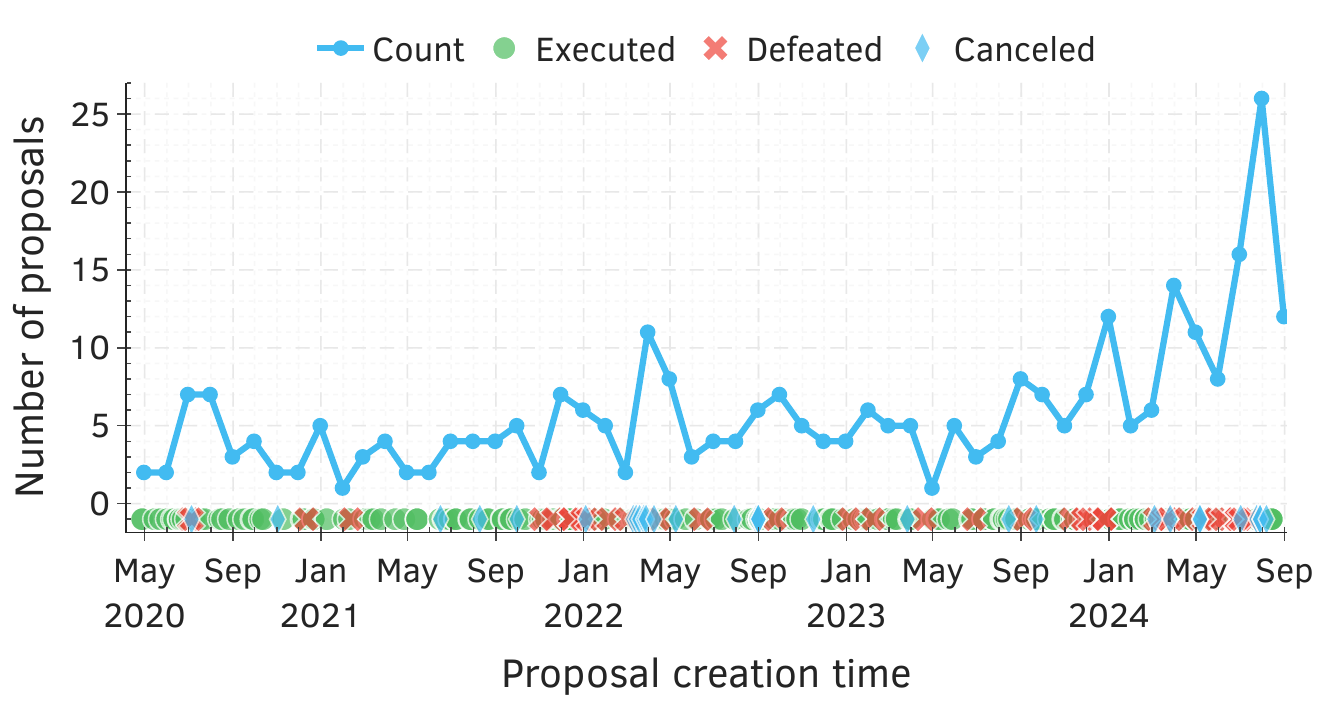}
        \caption{Compound}
        \label{fig:new_compound-proposals-month}
    \end{subfigure}
    \begin{subfigure}{\twocolgrid}
        \centering
        \includegraphics[width=\twocolgrid]{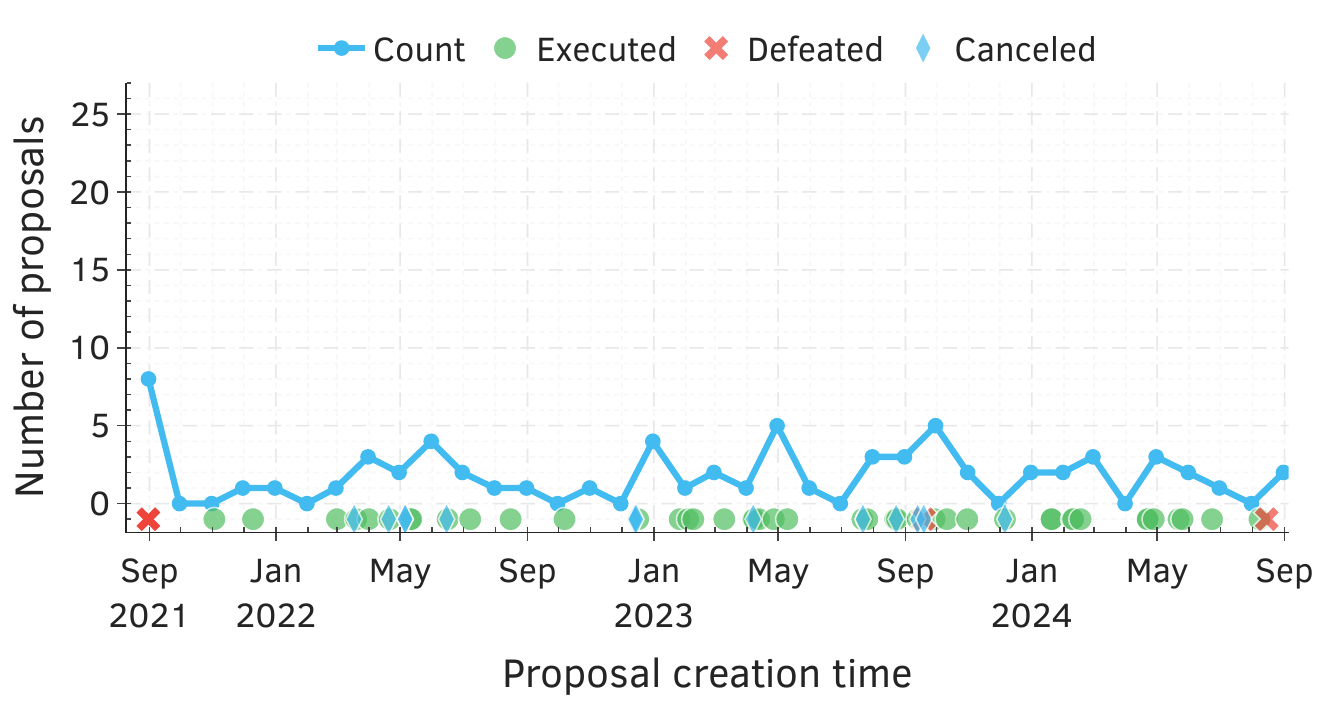}
        \caption{Uniswap}
        \label{fig:new_uniswap-proposals-month}
    \end{subfigure}
    \caption{
Monthly number of Compound and Uniswap proposals created and their outcome (\green{executed}, \red{defeated}, and \blue{canceled})}
    \label{fig:new-proposals-month}
\end{figure*}

Compound proposals are created frequently, with an average of \num{6.82} proposals per proposer. 
The highest number of proposals in Compound was created in July 2024, with \num{26} proposals (from \#270 to \#295). Of these, \num{20} were executed, \num{2} were defeated, and \num{4} were canceled (see Figure~\ref{fig:new_compound-proposals-month}). On average, proposals in Compound were submitted every \num{5.13} days (standard deviation \num{5.59}), with a median interval of \num{3.19} days. The shortest and longest intervals between proposals were \num{0} and \num{31.14} days, respectively. Proposals typically reached quorum in \num{1.69} days (standard deviation \num{0.69}) (Figure~\ref{fig:new_compound-time-to-quorum} in \S\ref{sec:compound-quorum}).

In comparison, Uniswap proposals are submitted less frequently. On average, one proposal is created every \num{16.36} days, with a standard deviation of \num{20.80} and a median interval of \num{6.48} days (Figure~\ref{fig:new_uniswap-proposals-month}). The longer intervals between proposals in Uniswap can likely be attributed to its more stable governance structure, which requires fewer updates compared to Compound’s more parameter-driven approach.

The shortest and longest intervals between proposals in Uniswap were \num{0} and \num{80.99} days, respectively. The largest number of proposals (\num{10}) occurred in August and September 2021. However, many of these proposals were initial submissions by the Uniswap Governor and did not receive votes or appear on the Uniswap website~\cite{Uniswap-Proposals-Page}. Upon further analysis, it was found that the ``target'' field in these proposals contained an Ethereum \stress{null address} (0x0), meaning that even if the proposals passed, no changes would have been made to any contracts. This likely serves as a precautionary measure to prevent conflicts with proposal numbering in the Governor Alpha contract, though this hypothesis remains unverified.

\begin{mdframed}[style=Takeaways]
\textbf{Takeaway:}
Compound sees a higher frequency of proposals when compared to Uniswap, which is likely due to the need for regular adjustments to the protocol's parameters.
\end{mdframed}

\subsubsection{Lifecycle of Proposals}\label{subsubsec:proposal_lifecycle}

\begin{figure}[tb]
\centering
\includegraphics[width=\onecolgrid]{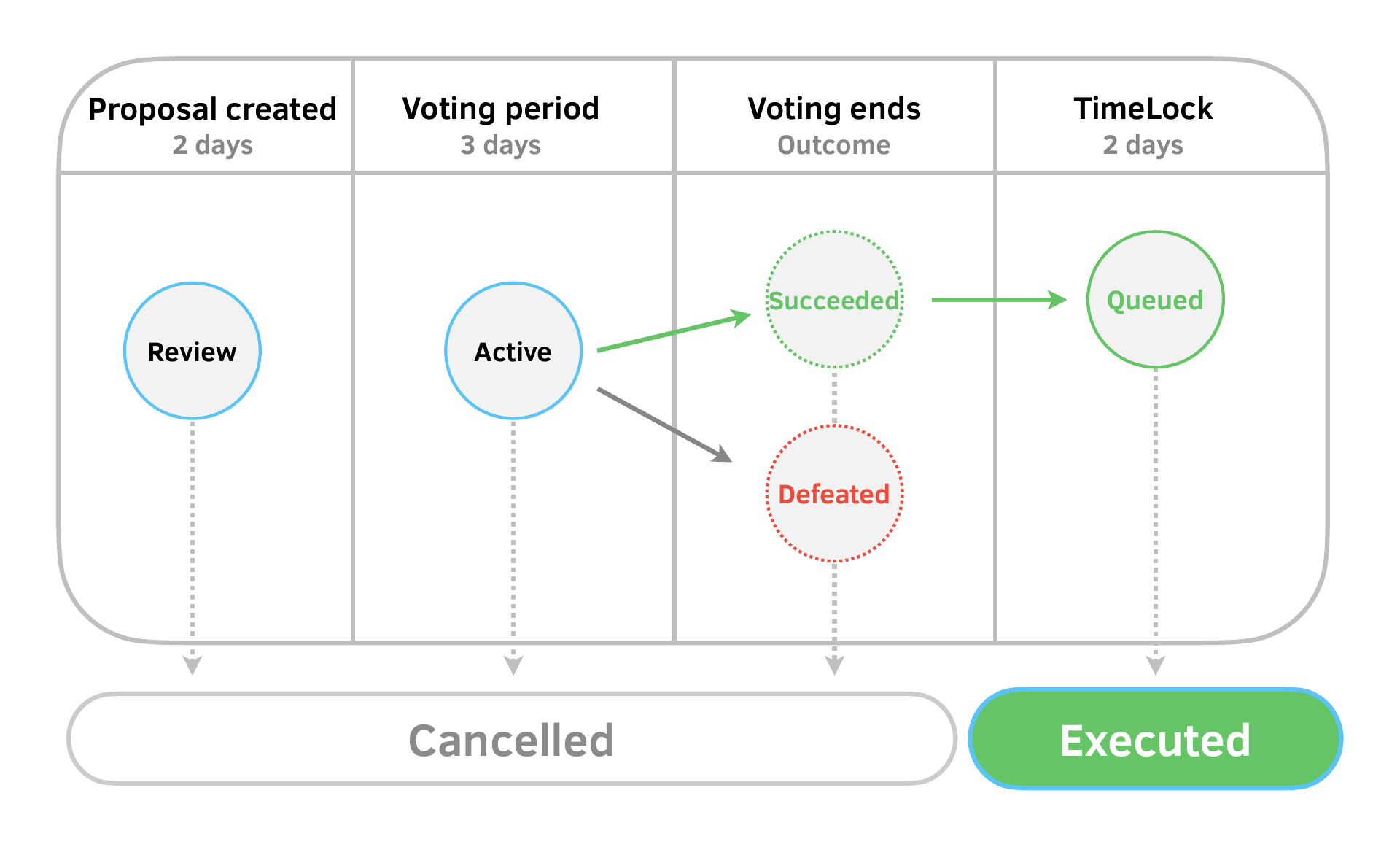}
\caption{ 
The lifecycle of a Compound proposal lasts 7 days. After a proposal is created, it waits for 2 days before the 3-day voting period begins. Once the outcome of the election is decided, it takes 2 more days for the proposal to be executed and become part of the Compound Governance protocol. Proposals can also be canceled at any time before they are executed.
}
\label{fig:compound-life-cycle}
\end{figure}

In both protocols, proposals undergo a voting period, where delegates cast their votes. Once a proposal receives the required quorum and majority, it is executed after a short time-lock period. The decision-making process for each protocol allows for governance decisions to be transparent and publicly available, but the effectiveness of these processes in decentralizing power depends on voter participation.

Per Figure~\ref{fig:compound-life-cycle}, when a Compound and Uniswap proposal is created, there is an approximately 2-day voting delay period (or \num{13140} blocks) that is used to allow the community to discuss the proposal before the voting period begins.
During the approximately 3-day (or \num{19710} blocks) for Compound and 7 days (or \num{40320} blocks) for Uniswap , voters can cast their votes. 
In order for a proposal to be executed, it needs to meet two requirements.
Firstly, it must receive a minimum of votes in favor of the proposal.
This number corresponds to \num{4}\% of the total supply in both Compound (\num{400000} tokens) and Uniswap (\num{40000000} tokens) and is known as the \stress{quorum}.
Secondly, the majority of the votes cast must be in favor of the proposal.
The number of votes each voter has is determined by the number of delegated tokens they held in the block before the voting period began.
This prevents voters from changing their delegated tokens after the voting has begun, which could potentially lead to sudden changes in the outcome of the election.
After a proposal is approved, it is placed in the \stress{TimeLock} for a minimum period of \num{2} days before it can be implemented. 
A proposal can be canceled at any time by the proposer prior to its execution, or by anyone if the proposer fails to maintain at least 0.25\% of the total supply delegated tokens.

\begin{mdframed}[style=Takeaways]
\textbf{Takeaway:}
Proposals require a 2-day discussion delay, followed by a voting period. It needs quorum (4\% of supply) and majority approval to pass, with voters' tokens fixed during voting. After approval, proposals face a minimum 2-day TimeLock before execution. Proposals can be canceled if the proposer fails to maintain 0.25\% of delegated tokens.
\end{mdframed}

\subsubsection{Proposal Outcome}\label{subsec:proposal-outcome}

\begin{figure*}[t]
    \centering
    \begin{subfigure}{\twocolgrid}
        \centering
        \includegraphics[width=\twocolgrid]{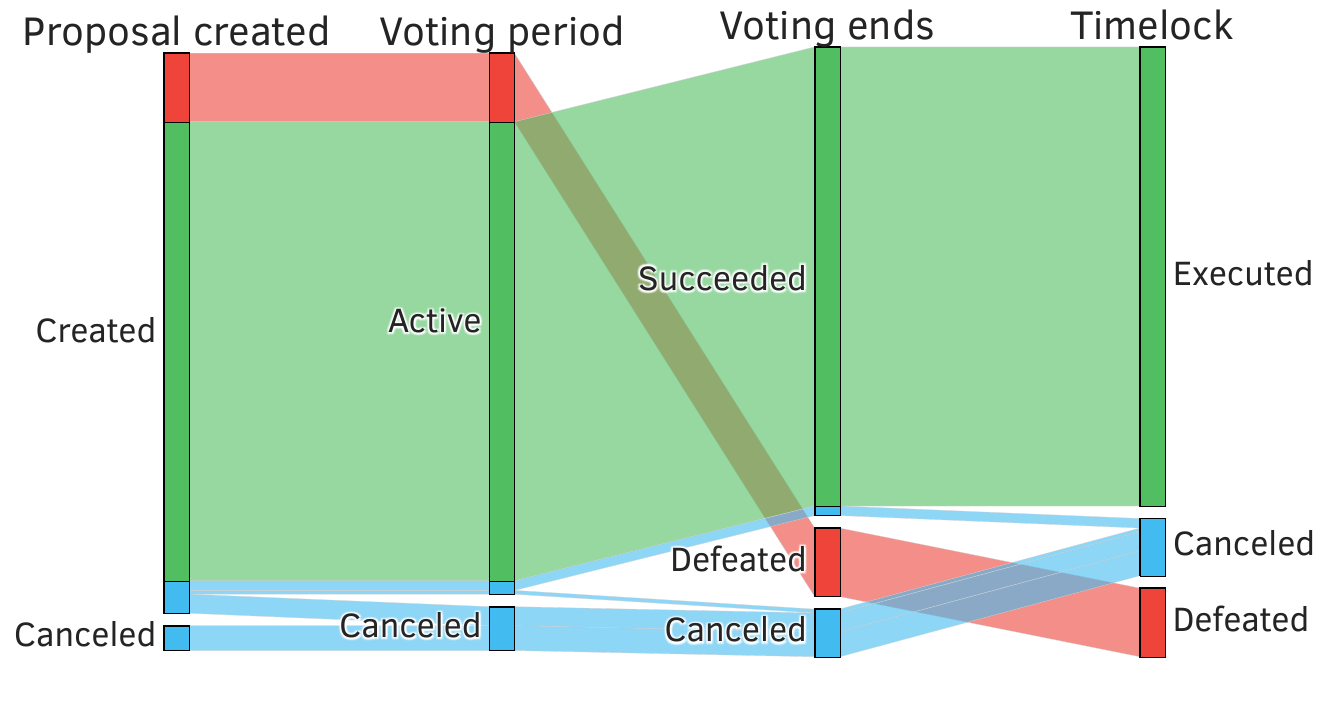}
        \caption{Compound}
        \label{fig:new_compound-proposal-life-cycle}
    \end{subfigure}
    \begin{subfigure}{\twocolgrid}
        \centering
        \includegraphics[width=\twocolgrid]{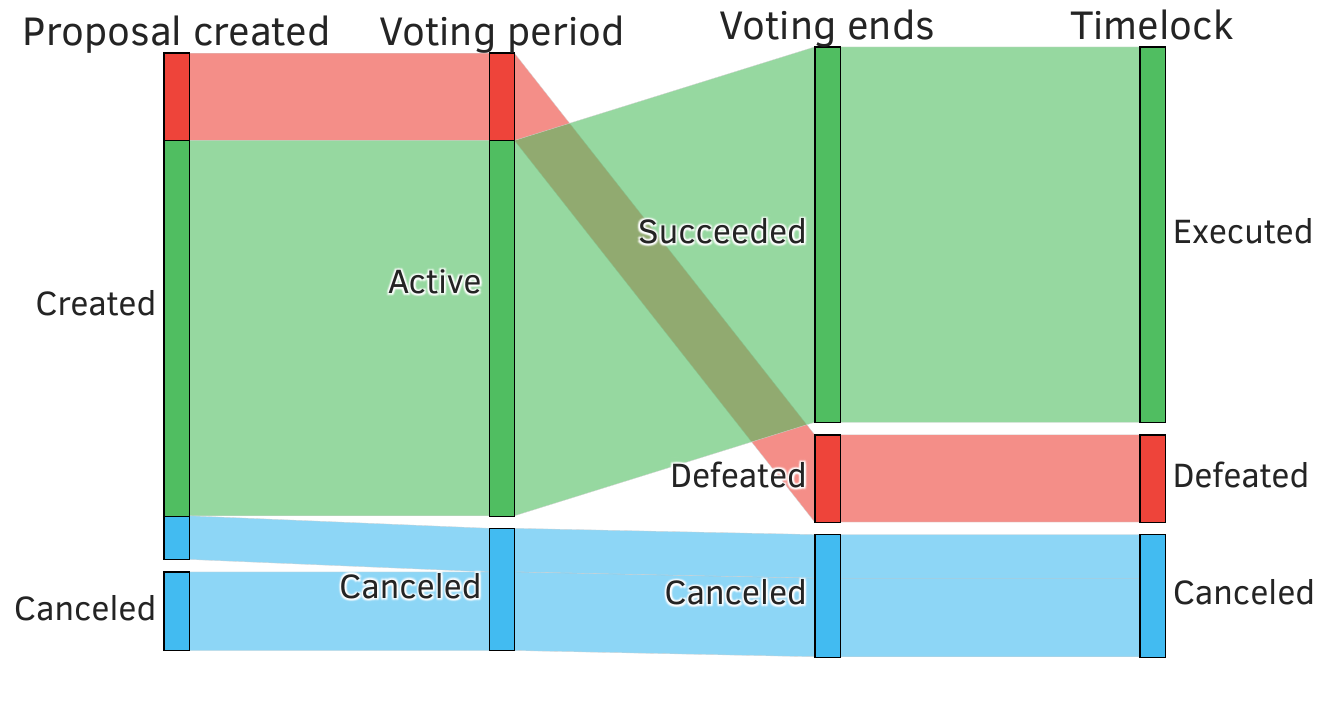}
        \caption{Uniswap}
        \label{fig:new_uniswap-proposal-life-cycle}
    \end{subfigure}\hfill
    \caption{
    Summary of the outcome of Compound and Uniswap proposals at each stage of their life cycle: (a) From 307 Compound proposals, 241 were executed (in \green{green}), 36 defeated (in \red{red}), and 30 canceled (in \blue{blue}); and (b) from 67 Uniswap proposals, 43 were executed, 10 defeated, and 14 canceled.
    }
    \label{fig:new-proposal-life-cycle}
\end{figure*}

During the analyzed period, \num{307} proposals were created in Compound. Of these, \num{30} were canceled and \num{36} were defeated, leaving \num{241} (\num{78.50}\%) proposals executed. The outcome of Compound’s proposals at each stage of their lifecycle is highlighted in Figure~\ref{fig:new_compound-proposal-life-cycle}. In comparison, for Uniswap, \num{67} proposals were analyzed, of which \num{43} (\num{64.18}\%) were executed, \num{10} (\num{14.93}\%) were defeated, and \num{14} (\num{20.90}\%) were canceled (refer to Figure~\ref{fig:new_uniswap-proposal-life-cycle}). All defeated proposals on Uniswap failed due to not reaching quorum. In contrast, all but one of the defeated proposals on Compound were rejected for not reaching quorum. This suggests that users may be less inclined to participate in voting when a proposal is unlikely to reach quorum, potentially timing their participation accordingly. We provide further details in Section~\ref{subsec:timing-voting}.

Our analysis further shows that \num{13} Compound proposals and \num{9}  Uniswap proposals were canceled immediately after creation, meaning they never reached the \stress{Voting Period} and were not available for voting. Additionally, \num{10} Compound proposals and \num{5} Uniswap proposals were canceled before the \stress{Voting Ends} stage, meaning they were withdrawn before the election was concluded.

Surprisingly, \num{2} Compound proposals were canceled after they passed the vote but before being queued in the \stress{Timelock}. Another \num{5} Compound proposals were canceled while in the \stress{Timelock}, which may reflect a lack of community consensus~\cite{sharma2023unpacking}.

\begin{mdframed}[style=Takeaways]
\textbf{Takeaway:}
The majority of proposals were executed on both protocols.
Some proposals were canceled before or during voting, including 2 in Compound after passing the vote but before TimeLock, and 5 during TimeLock, possibly due to community consensus issues.
\end{mdframed}

\subsection{Dynamics of Governance Voting}

This section analyzes how governance proposals are actually voted on, focusing on proposal support, voter participation, voting costs, and timing strategies. We uncover key dynamics such as low engagement, cost barriers, and strategic voting that challenge the fairness and effectiveness of decentralized governance in practice.

\subsubsection{Support and Participation Trends}

\begin{figure*}[t]
    \centering
    \begin{subfigure}{\twocolgrid}
        \centering
        \includegraphics[width=\twocolgrid]{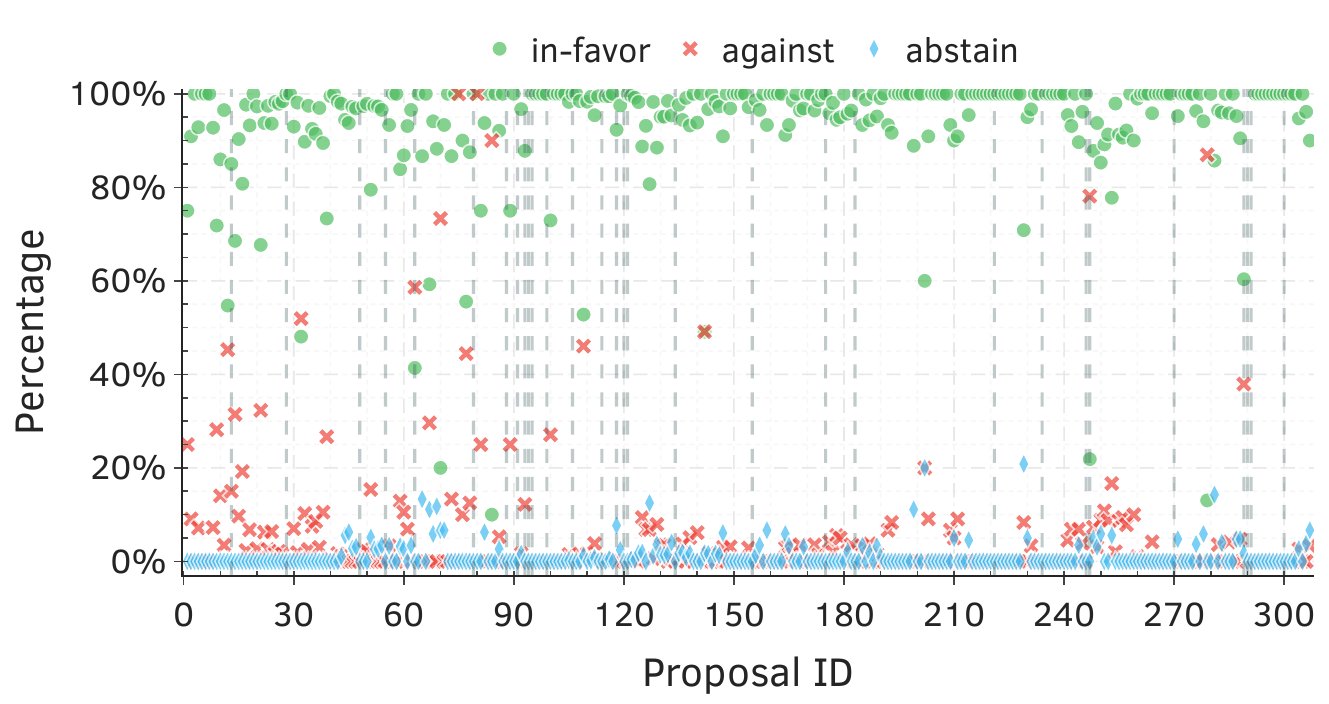}
        \caption{Compound}
        \label{fig:new_compound-votes-proposal-percentage}
    \end{subfigure}
    \begin{subfigure}{\twocolgrid}
        \centering
        \includegraphics[width=\twocolgrid]{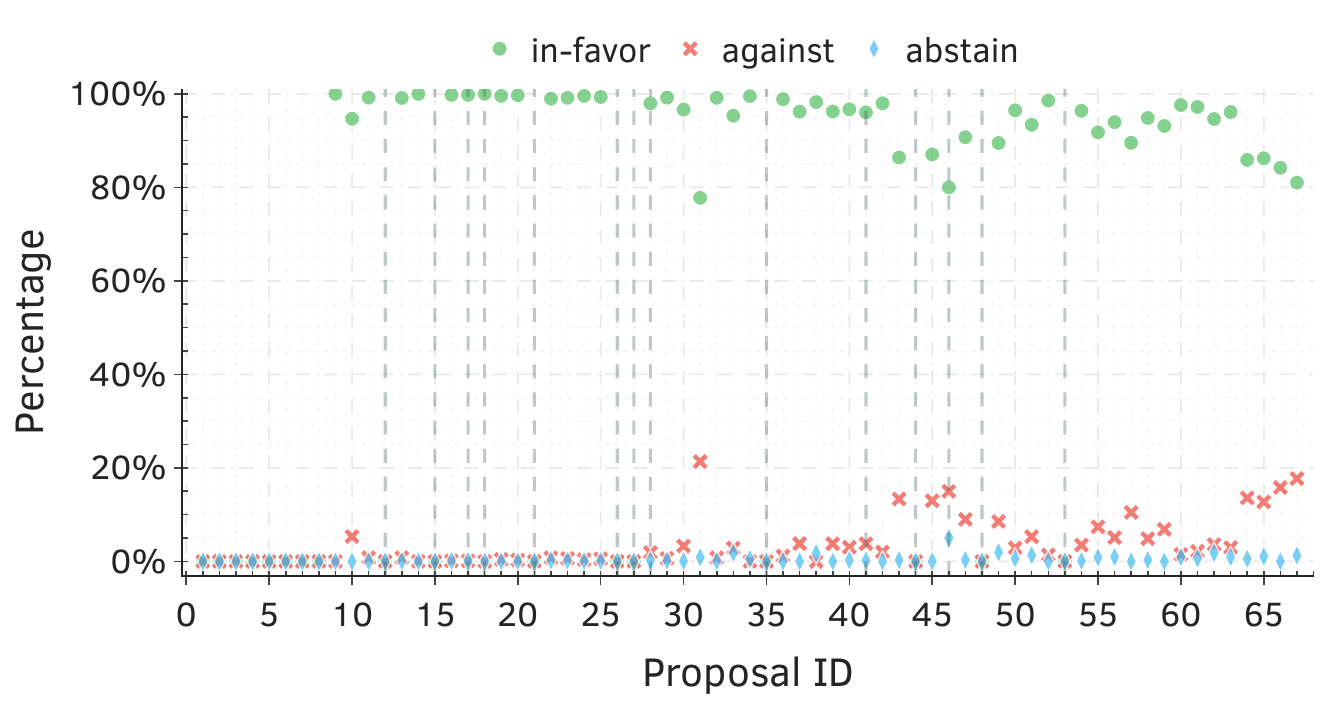}
        \caption{Uniswap}
        \label{fig:new_uniswap-votes-proposal-percentage}
    \end{subfigure}\hfill
    \caption{
    Percentage of in-favor (in \green{green}), against (in \red{red}), and abstain (in \blue{blue}) votes for each Compound and Uniswap proposals. Vertical lines indicate canceled proposals.
    }
    \label{fig:new-votes-proposal-percentage}
\end{figure*}

\begin{figure*}[tb]
\centering
\includegraphics[width=\onecolgrid]{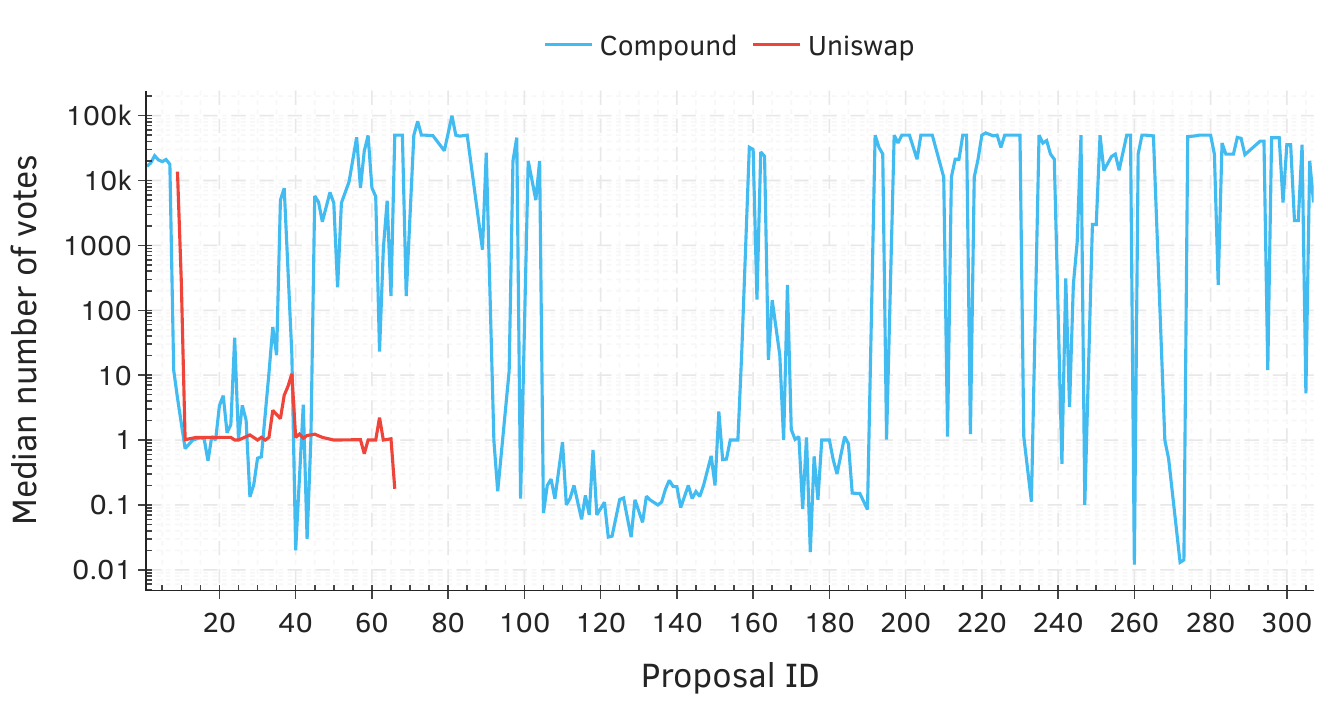}
\caption{Distribution of voting power by voter per proposal for Compound and Uniswap. We consider a cutoff of 0.001 votes.}
\label{fig:voting-distribution-per-proposal}
\end{figure*}

Before proposals are officially submitted, the community often conducts a ``temperature check'' to gauge support for the proposal (see \S\ref{subsubsec:proposal_lifecycle}). This pre-submission engagement suggests that proposals are likely to receive significant support from voters.

Our analysis, shown in Figure~\ref{fig:new_compound-votes-proposal-percentage}, confirms this trend for Compound. The figure displays the percentage of votes cast in favor, against, and abstaining for each proposal. Most proposals received strong support, with an average of \num{88.63}\% of votes in favor, a standard deviation of \num{24.49}\%, and a median of \num{97.66}\%.
A similar pattern is observed for Uniswap (Figure~\ref{fig:new_uniswap-votes-proposal-percentage}). Proposals in Uniswap passed with an average of \num{70.73}\% of votes in favor and a median of \num{94.67}\%.

Next, we analyze the votes cast in both Compound and Uniswap. In total, \num{14841} votes were cast in Compound, and \num{51580} votes were cast in Uniswap. Of these, \num{2463} (\num{16.60}\%) of the votes in Compound came from voters who did not have any delegated tokens, resulting in a ``useless vote'' with zero voting power. In Uniswap, this occurred for \num{4168} votes (\num{8.08}\%). While such votes do not count for or against a proposal, they still show support for the proposal, as these voters are participating in the election despite not having voting power.

In Compound, the average number of votes cast per proposal (or tokens used to vote) was \num{13277.57}, with a standard deviation of \num{40143.13} and a median of \num{0.15}, ranging from \num{0} to \num{361006.43}. In Uniswap, the number of votes ranged from \num{0} to \num{15019646.45}, with an average of \num{47548.76} votes, a standard deviation of \num{491462.89}, and a median of \num{1}. This median indicates that most voters in both Compound and Uniswap hold relatively small amounts of voting power (see Figure~\ref{fig:voting-distribution-per-proposal}).

Notably, we observed four highly disputed proposals in Compound with narrow voting margins, less than 10\% difference between votes in favor and against, indicating a higher degree of contention. Among these, one proposal was defeated (ID \#13), two were canceled (IDs \#13 and \#289), and only one was successfully executed (ID \#16) (see Figure~\ref{fig:new_compound-votes-proposal-percentage}). In contrast, we did not observe any disputed proposals in Uniswap.

\begin{mdframed}[style=Takeaways] \takeaways{Takeaway:}
Pre-submission engagement boosts proposal support, but voting power remains concentrated among a few voters. Some votes are cast by participants without delegated tokens (i.e., voting power).
\end{mdframed}

\subsubsection{Voter Engagement Rates}

\begin{figure*}[t]
    \centering
    \begin{subfigure}{\twocolgrid}
        \centering
        \includegraphics[width=\twocolgrid]{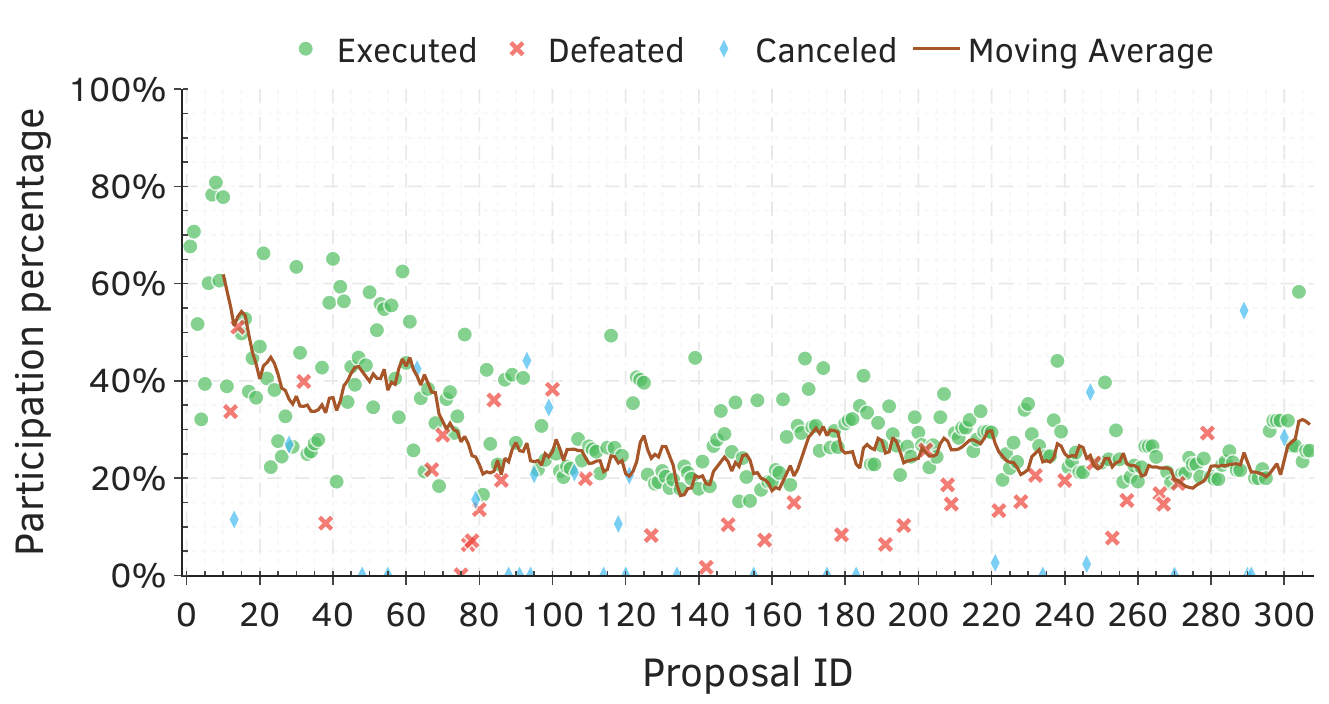}
        \caption{Compound}
        \label{fig:new_comp-voting-participation}
    \end{subfigure}
    \begin{subfigure}{\twocolgrid}
        \centering
        \includegraphics[width=\twocolgrid]{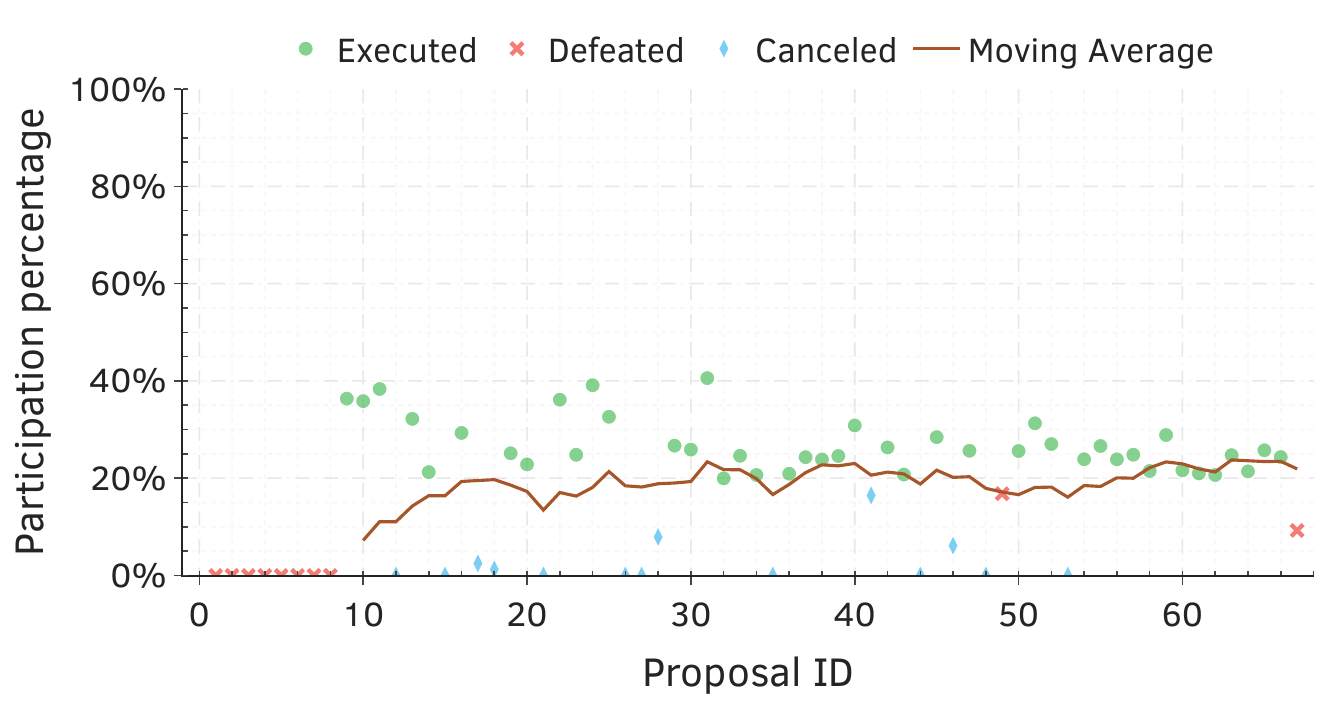}
        \caption{Uniswap}
        \label{fig:new_uni-voting-participation}
    \end{subfigure}\hfill
    \caption{
    Compound's and Uniswap's voting participation per proposal in terms of delegated tokens used from all delegated tokens available.}
    \label{fig:new_voting-participation}
\end{figure*}

The voter participation rate is a key indicator of community engagement in the governance process. Therefore, we calculated the voting participation per proposal (see Figure~\ref{fig:new_voting-participation}). This metric is computed by dividing the number of votes (or delegated tokens) cast on a proposal by the total number of delegated tokens eligible to vote at the start of the voting period. This measure is crucial as it reflects the proportion of delegated tokens used in the governance process per proposal. Low voter turnout can make governance systems more susceptible to \stress{vote-buying}, as non-voting token holders may sell their voting rights to others~\cite{Daian-dark-dao@HackingDistributed}, making such power concentration even worse.

Our results show that in Compound, the average voter turnout is \num{28.03}\% (Figure~\ref{fig:new_comp-voting-participation}), with a maximum of \num{80.80}\%. In Uniswap, the average turnout is \num{18.08}\% (Figure~\ref{fig:new_uni-voting-participation}), with a maximum of \num{40.58}\%. These relatively low participation rates suggest that many token holders are not actively engaging in the voting process, which can undermine the effectiveness of the governance system.

As shown in Figure~\ref{fig:new_comp-voting-participation}, voting participation in Compound was higher for earlier proposals, likely due to the limited availability of tokens at the outset. On average, \num{50.65} voters participated in each of the \num{307} proposals, with a standard deviation of \num{71.72} and a median of \num{29} voters. Voter participation ranged from \num{0} (for canceled proposals) to a maximum of \num{619} voters, as seen in proposal \#111, which garnered \num{686289.04} votes from \num{615} voters in favor, \num{3} against, and \num{1} abstaining.
In comparison, Uniswap (Figure~\ref{fig:new_uni-voting-participation}) had a higher average number of voters per proposal, with \num{1031.60} voters on average (standard deviation of \num{1131.84}) and a median of \num{686.50} voters. The range of voters per proposal in Uniswap varied from \num{0} to \num{5836}.

\begin{mdframed}[style=Takeaways]
\textbf{Takeaway:} Low voter participation rates in both Compound and Uniswap raise concerns about the effectiveness of decentralized governance, as a large proportion of token holders do not participate in decision-making. However, Uniswap token holders participate more when compared to Compound's.
\end{mdframed}

\subsubsection{Voting Costs Analysis}

\begin{figure*}[t]
    \centering
    \begin{subfigure}{\twocolgrid}
        \centering
        \includegraphics[width=\twocolgrid]{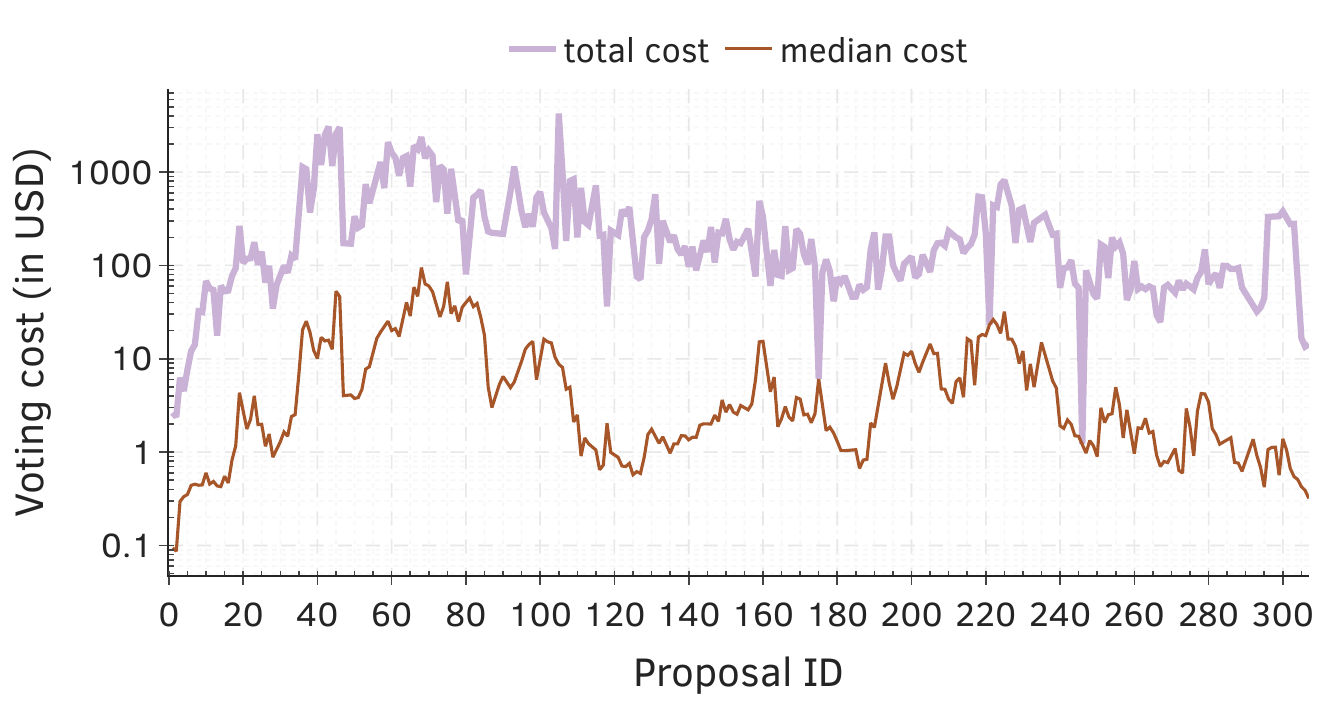}
        \caption{Compound}
        \label{fig:new_compound-voting-cost-per-proposal}
    \end{subfigure}
    \begin{subfigure}{\twocolgrid}
        \centering
        \includegraphics[width=\twocolgrid]{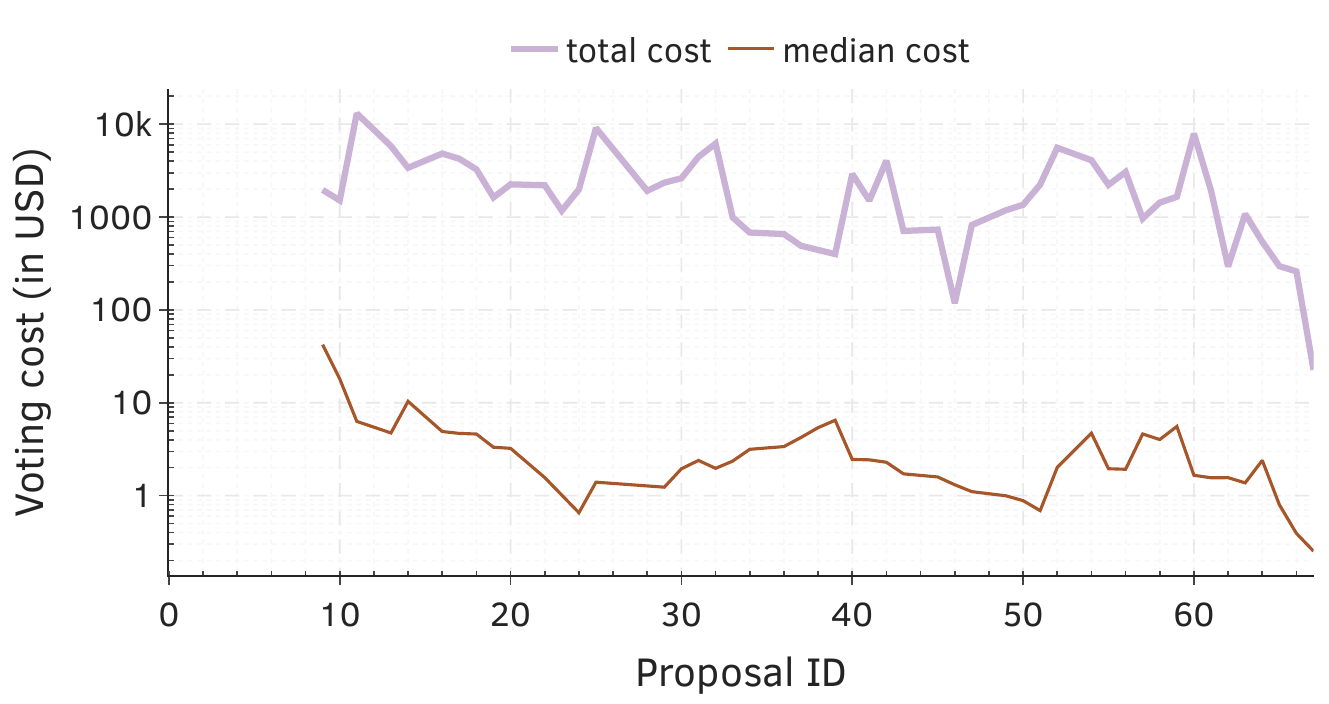}
        \caption{Uniswap}
        \label{fig:new_uniswap-voting-cost-per-proposal}
    \end{subfigure}
    \caption{
    Voting cost distribution per proposal: (a) Compound, on average votes cost \$6.82 with a std. of \$19.04; and (b) Uniswap, it costed an average of \$2.42 with a std. of \$2.56. 
    }
    \label{fig:new_voting-cost-per-proposal}
\end{figure*}

Voting costs can act as a deterrent to participation, particularly for smaller token holders. These costs are influenced by Ethereum network transaction fees, which fluctuate based on network congestion. To assess the impact of voting costs, we analyzed relevant transactions from the Ethereum blockchain, focusing on the fees paid by voters to cast their votes. We report the voting costs in US dollars, using the ETH-USD exchange rate from the Yahoo Finance data feed~\cite{eth-usd@yahoo} at the time the transaction was processed.

Across all \num{307} Compound proposals, the total voting cost amounted to \$\num{101154.27}, while for Uniswap, the total was \$\num{124571.51}, approximately 1.23 times higher. In Compound, voting costs ranged from \$0.03 to \$294.02, with an average cost of \$6.82 and a median of \$1.67 (see Figure~\ref{fig:new_compound-voting-cost-per-proposal}). In Uniswap, voting costs ranged from \$0.17 to \$126.86, with an average of \$2.42 and a median of \$1.83 (see Figure~\ref{fig:new_uniswap-voting-cost-per-proposal}).

To gain a deeper understanding, we normalized the cost of casting a vote by the number of votes cast, measured by the total number of delegated tokens available to each voter's address.  We found that some voters faced disproportionately high costs per vote unit. For example, the mean cost per vote unit in Compound was \$\num{598.97} (with a standard deviation of \$\num{7739.34}), indicating a highly skewed distribution. However, half of the voters faced a cost per vote unit of just \$\num{6.38}. In Uniswap, the mean cost per vote unit was \$\num{102.56} (with a standard deviation of \$\num{2814.79}), and the median cost was \$\num{1.75}.

\begin{mdframed}[style=Takeaways]
\takeaways{Takeaway:} The high cost of voting disproportionately affects smaller stakeholders, creating a financial barrier that limits their participation and undermines the principle of fair governance.
\end{mdframed}

\subsection{Impact of Voting Gaps}\label{subsec:timing-voting}

\begin{figure*}[tb]
\centering
\begin{subfigure}{\twocolgrid}
  \centering
  \includegraphics[width=\twocolgrid]{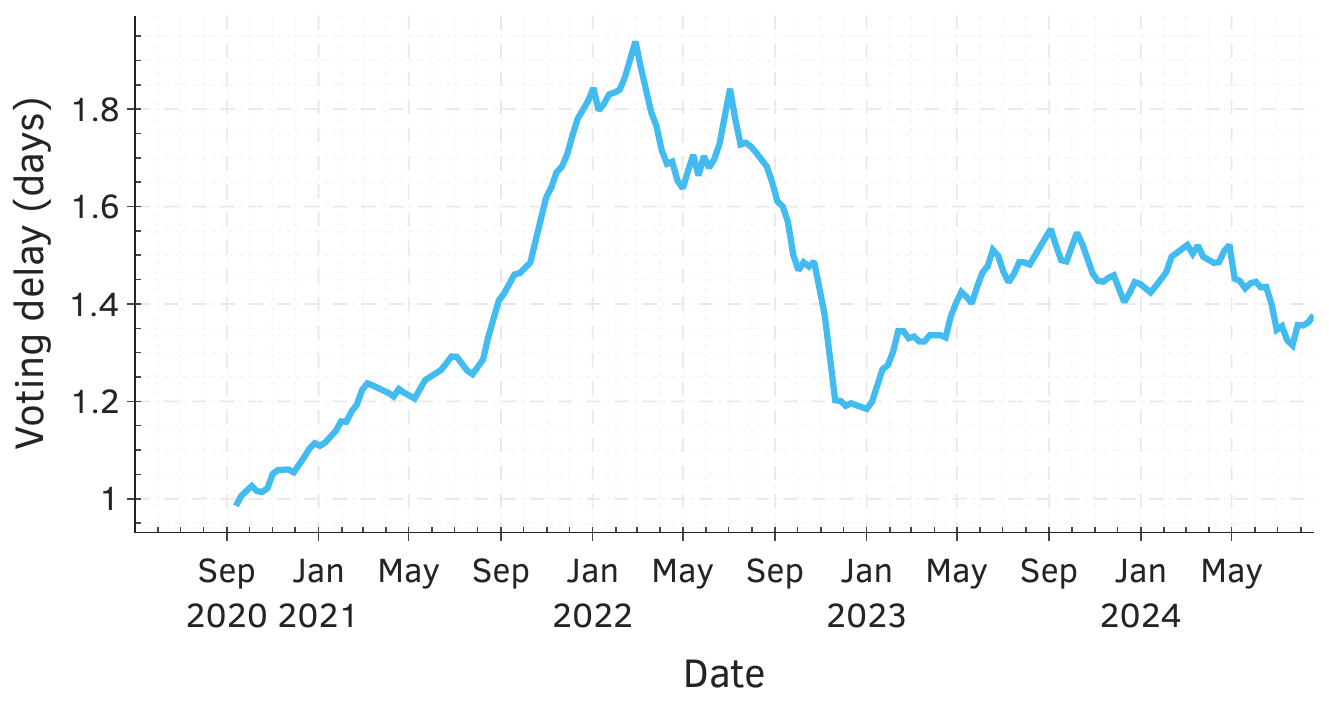}
  \caption{Compound}
  \label{fig:avg_vote_delay_per_day_compound}
\end{subfigure}
\begin{subfigure}{\twocolgrid}
  \centering
  \includegraphics[width=\twocolgrid]{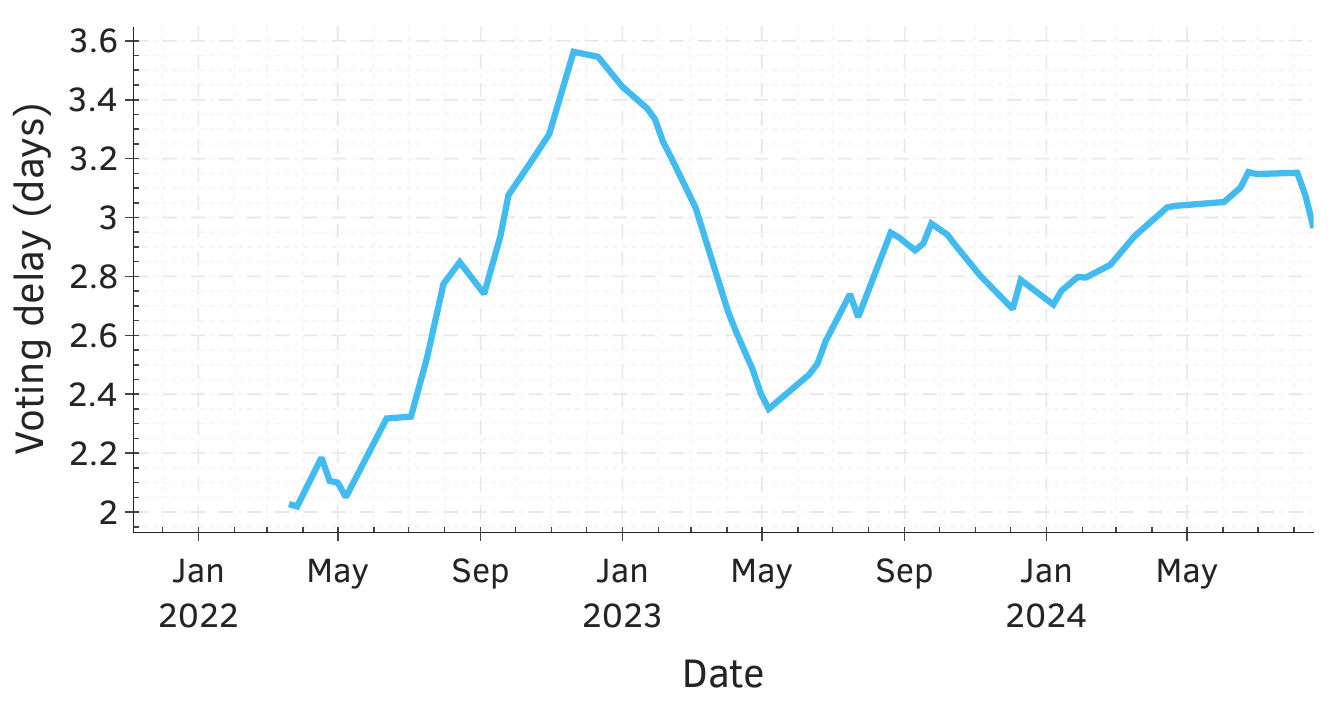}
  \caption{Uniswap}
  \label{fig:avg_vote_delay_per_day_uniswap}
\end{subfigure}
\caption{Moving average of the voting delay (in days) relative to the start of the voting period.}
\label{fig:avg_vote_delay_per_day}
\end{figure*}

Casting a vote on-chain incurs transaction costs, which can discourage participation, especially when a proposal appears unlikely to reach quorum, as observed in Section~\ref{subsec:proposal-outcome}. In such cases, rational voters may choose to abstain, avoiding unnecessary expenses.

Governance protocols like Compound and Uniswap do not allow vote changes once a vote is cast; each participant can vote only once per proposal. Despite this restriction, the transparency of on-chain voting may enable strategic timing, where voters delay their participation to assess the likely outcome before committing. This behavior is captured by the concept of the voting gap, the time elapsed between the start of the voting period and when a vote is submitted. Strategic delays, also referred to as voting sniping~\cite{feichtinger2024sok,yaish2024strategic}, can significantly influence proposal outcomes, particularly in contentious cases.

Our empirical analysis reveals that Compound voters cast their votes after an average delay of \num{1.43} days (standard deviation \num{0.90}, median \num{1.38}), with delays ranging from \num{0} to \num{3.39} days. In Uniswap, the average delay is notably longer at \num{2.72} days (standard deviation \num{1.71}, median \num{2.55}), with a maximum observed delay of \num{7.27} days. The moving average of these voting gaps is shown in Figure~\ref{fig:avg_vote_delay_per_day}.

In parallel, we find that proposals typically reach quorum in \num{1.69} days (standard deviation \num{0.69}), as shown in Figure~\ref{fig:new_compound-time-to-quorum} in \S~\ref{sec:compound-quorum}. As discussed in Section~\ref{subsec:proposal-outcome}, nearly all Uniswap proposals and all but one Compound proposal failed due to not meeting quorum thresholds. This further supports the hypothesis that voters strategically withhold participation when a proposal appears unlikely to succeed, thereby conserving gas fees.

\begin{mdframed}[style=Takeaways]
\takeaways{Takeaway:}
On-chain voting behavior is shaped by transaction costs and proposal viability. Voters often delay or abstain from voting when outcomes seem predetermined, enabling strategic timing that can distort the decision-making process and hinder quorum formation. 
\end{mdframed}

\section{Discussion}
\label{sec:discussion}

This section discusses the broader implications of our findings on decentralized governance protocols. We also outline potential future research directions to enhance governance protocols and better align them with the principles of decentralization and fairness.

\subsection{Fairness and Centralization Concerns}
\label{sec:fairness-decentralization}

One of the key concerns identified in our study is the concentration of voting power within a small group of participants, which raises significant issues for the fairness and decentralization of governance protocols. Our analysis shows that in both Compound and Uniswap, a small number of voters control a disproportionate amount of voting power. For instance, 10 voters controlled \num{50.53}\% of the voting power in Compound, while the top 10 voters in Uniswap controlled \num{35.73}\% of the total voting power. This level of concentration is concerning and undermines the ideal of decentralized decision-making.

In this sense, the concentration of power can result in governance decisions that do not reflect the preferences of the broader community. While delegated voting mechanisms allow for more efficient decision-making, they can also lead to the centralization of power, particularly when a few large stakeholders delegate their votes to unknown representatives. This creates an imbalance in which the most powerful participants dominate the decision-making process, which contradicts the core principles of decentralization and fairness. In practice, governance protocols may not be as decentralized as they appear, with decisions being disproportionately influenced by a small number of participants.

\subsection{Voting Cost Barriers}
\label{sec:cost-barriers}

Our findings reveal that the cost of participating in governance can be a significant barrier, particularly for smaller stakeholders. In both Compound and Uniswap, we observed a wide range of transaction fees for voting, from as low as \$\num{0.03} to as high as \$\num{294.02} in Compound, and \$\num{0.17} to \$\num{126.86} in Uniswap. These fees create a barrier to participation, particularly for smaller stakeholders who may not have the financial resources to engage in the voting process. This creates a ``poll tax'' effect, where only those with substantial holdings can afford to vote, undermining the principle of equal participation in governance.

Moreover, when we normalize the voting costs by the number of tokens held, smaller stakeholders face disproportionately high costs relative to their economic stake in the protocol. This disparity creates an inequitable system where the governance process is dominated by larger stakeholders, reinforcing centralization and reducing the fairness of decision-making. To ensure that decentralized governance remains inclusive and representative, it is crucial to reduce voting costs and make the process more accessible to all participants, particularly those with smaller stakes.

\subsection{Future Research Directions}
\label{sec:future-research}

There are several avenues for future research aimed at improving decentralized governance systems. One potential direction is to explore alternative voting mechanisms that can reduce the concentration of voting power, such as quadratic voting~\cite{Lalley@AEA}. Quadratic voting, for instance, allows participants to allocate more votes to proposals they care about, but the cost of additional votes increases quadratically, which helps to reduce the influence of large stakeholders. However, Sybil resistance might be considered carefully.

Another area for future research is the development of more efficient and cost-effective voting systems that reduce the transaction fees associated with voting. One possibility is the introduction of gas subsidies or the use of layer-2 scaling solutions to reduce costs, making voting more accessible to smaller stakeholders.

Moreover, another way to mitigate the concentration of voting power would be to improve platforms used to delegate tokens. These platforms provide dashboards and websites that display critical information about DAOs, including the distribution of delegated tokens, historical voting records, and proposal submissions. However, the way these platforms rank accounts can inadvertently bias token holders toward delegating to the highest-ranked participants. Such a bias may lead to a \textit{rich get richer} dynamic~\cite{Rich-gets-Richer@Wikimedia}, concentrating voting power and potentially undermining the decentralized promise of these protocols. Popular platforms like Tally are already in use by leading projects in the crypto ecosystem, including Arbitrum~\cite{Arbitrum@Tally}, Wormhole~\cite{Wormhole@Tally}, Compound~\cite{Compound@Tally}, Uniswap~\cite{Uniswap@Tally}, ENS~\cite{ENS@Tally}, and ZKsync~\cite{ZKsync@Tally}.

Additionally, further work is needed to investigate the long-term implications of strategic voting behaviors, such as voting delays or ``voting sniping.'' Understanding the impact of such strategies on governance outcomes and exploring mechanisms to mitigate their effects will be crucial for ensuring the integrity of decentralized decision-making.

\section{Related Work}\label{sec:related}

The literature on decentralized governance, social contracts, and decentralized autonomous organizations (DAOs) is extensive.

\paraib{Decentralized Governance and Social Contracts}
Prior work has investigated how blockchain-based governance might serve as an alternative to traditional centralized systems. Atzori~\etal{} examined the potential for blockchain to displace centralized societal institutions and authorities~\cite{atzori2017blockchain}. Reijers~\etal{} analyzed the alignment between blockchain governance and classical theories of the social contract, drawing on the ideas of Hobbes~\cite{hobbesleviathan}, Rousseau~\cite{rousseau1920social}, and Rawls~\cite{john1971rawls}. Chen~\etal{} explored trade-offs between decentralization and system performance~\cite{chen2021decentralized}. Other comparative studies have examined governance models across public and private blockchains~\cite{arrunada2018blockchain,zwitter2020decentralized}. For instance, Arruñada and Garicano proposed new forms of “soft” decentralized governance~\cite{arrunada2018blockchain}, while Zwitter and Hazenberg offered a rethinking of governance structures tailored to DAOs~\cite{zwitter2020decentralized}. While these works provide conceptual insights, they generally lack empirical analysis of how these theories play out in operational DAO systems or how real-world governance protocols function in practice.

\paraib{Decentralized Autonomous Organizations (DAOs)}
Several studies have examined the internal governance structures of DAOs~\cite{beck2018governance,dotan2023vulnerable,hassan2021decentralized,rikken2019governance}. Hassan and De Filippi investigated the fundamental properties and constitutional frameworks of DAOs~\cite{hassan2021decentralized}, while Rikken~\etal{} identified political challenges that arise within blockchain-based governance systems~\cite{rikken2019governance}. Beck~\etal{} presented a case study of Swarm City, a decentralized platform for commerce~\cite{beck2018governance,swarm-city}. More recent work has sought to categorize governance structures across blockchains and protocols, including Bitcoin, Ethereum, Tezos, and Polkadot, as well as major DeFi protocols such as Uniswap~\cite{adams2021uniswap}, MakerDAO~\cite{Governance@MakerDAO}, and Compound~\cite{leshner2019compound,kiayias@2022governance}. While these contributions outline the architecture of DAO governance, most stop short of empirically analyzing user interactions with governance mechanisms or measuring their practical implications.

Moreover, recent efforts have begun addressing DAO vulnerabilities and the measurement of decentralization. Feichtinger~\etal{} compiled a systematization of attacks on DAOs, categorizing them by their attack vectors and offering insight into the risks facing decentralized governance~\cite{feichtinger2024sok}. Austgen~\etal{} introduced a framework for evaluating DAO decentralization through factors such as herding and delegation, and proposed a new metric called Voting-Bloc Entropy (VBE)~\cite{austgen2023dao}. However, while theoretically grounded, their analysis lacks empirical validation on how these factors manifest in practice. Tan~\etal{} compiled a comprehensive set of open research questions surrounding DAOs across disciplines including computer science, economics, and law~\cite{tan2023open}. Although valuable, their work does not examine how DAOs are used in real-world scenarios or the implications of their governance in practice.

\paraib{Empirical Studies Closely Related to This Work}
Three lines of empirical research are especially relevant to our study~\cite{feichtinger2023hidden,kitzler2023governance,sharma2023unpacking}. These studies report significant centralization in DAO governance: a high concentration of delegated voting power and greater participation by large token holders. While these works analyze governance across multiple DAOs, our study offers a focused and deeper examination of two prominent protocols, Compound and Uniswap.
Our contribution extends previous findings in several directions. First, we analyze token concentration, voting delays, coalition formation, and the unequal burden of participation among small vs. large voters. Second, we examine the full lifecycle of proposals to understand how voting behavior evolves over time. Third, by using labeled accounts rather than raw blockchain addresses, we provide a more nuanced view of user behavior, including the identification of voter coalitions and patterns of strategic delegation. Finally, we expose disparities in voting costs that disproportionately affect smaller participants, highlighting critical barriers to inclusive governance in practice.

\section{Conclusion}
\label{sec:conclusion}

This study provides the first in-depth comprehensive empirical analysis of the governance practices in two foundational DeFi protocols: Compound and Uniswap. Our findings reveal a stark discrepancy between the decentralized ideals of these protocols and their actual governance outcomes. In both protocols, we observe extreme concentration of voting power among a handful of participants, with just 3.18 voters in Compound and 4.7 in Uniswap typically sufficient to achieve majority control. These dynamics are exacerbated by voting cost asymmetries, where smaller stakeholders bear disproportionately high transaction costs, effectively pricing them out of governance participation.

Additionally, we detect strategic voting behaviors, including delayed participation and coalition-based voting patterns, that may manipulate proposal outcomes and undermine collective decision-making. Such findings point to systemic issues in DAO design: \stress{while technically decentralized, governance remains vulnerable to economic centralization and behavioral manipulation.}

To address these shortcomings, we advocate for more equitable voting mechanisms such as quadratic voting to rebalance influence, the integration of cost-reduction strategies like gas subsidies or Layer-2 solutions to lower participation barriers, and more transparent delegation tools to counteract herd behavior.
Finally, we make all code and data used in our study publicly available to support reproducibility and encourage further research into improving the integrity and inclusiveness of decentralized governance~\cite{Messias-Dataset-Code-2025}.


\bibliographystyle{plainurl}
\bibliography{references}

\begin{thebibliography}{10}

\bibitem{adams2021uniswap}
Hayden Adams, Noah Zinsmeister, Moody Salem, River Keefer, and Dan Robinson.
\newblock {Uniswap v3 core}, 2021.

\bibitem{Governance@a16z}
Jeff Amico.
\newblock On crypto governance.
\newblock \url{https://a16z.com/2021/02/05/on-crypto-governance}, 2023.
\newblock Accessed on February 8, 2023.

\bibitem{arrunada2018blockchain}
Benito Arru{\~n}ada and Luis Garicano.
\newblock Blockchain: The birth of decentralized governance.
\newblock {\em Pompeu Fabra University, Economics and Business Working Paper Series}, 1608, 2018.

\bibitem{atzori2017blockchain}
Marcella Atzori.
\newblock Blockchain technology and decentralized governance: Is the state still necessary?
\newblock {\em Journal of Governance and Regulation/Volume}, 6(1), 2017.

\bibitem{austgen2023dao}
James Austgen, Andr{\'e}s F{\'a}brega, Sarah Allen, Kushal Babel, Mahimna Kelkar, and Ari Juels.
\newblock Dao decentralization: Voting-bloc entropy, bribery, and dark daos.
\newblock {\em arXiv preprint arXiv:2311.03530}, 2023.

\bibitem{Governance@Balancer}
{Balancer.fi}.
\newblock {Governance -- Balancer}.
\newblock \url{https://docs.balancer.fi/ecosystem/governance}, 2023.
\newblock Accessed on May 25, 2023.

\bibitem{beck2018governance}
Roman Beck, Christoph M{\"u}ller-Bloch, and John~Leslie King.
\newblock Governance in the blockchain economy: A framework and research agenda.
\newblock {\em Journal of the Association for Information Systems}, 19(10), 2018.

\bibitem{behrens2017origins}
Jan Behrens.
\newblock The origins of liquid democracy.
\newblock {\em The Liquid Democracy Journal}, 5(2), 2017.

\bibitem{Bluesky}
{Bluesky}.
\newblock {Get-started}.
\newblock \url{https://docs.bsky.app/docs/get-started}, 2024.
\newblock Accessed on May 15, 2024.

\bibitem{blum2016liquid}
Christian Blum and Christina~Isabel Zuber.
\newblock Liquid democracy: Potentials, problems, and perspectives.
\newblock {\em Journal of political philosophy}, 24(2), 2016.

\bibitem{carroll1884principles}
Lewis Carroll.
\newblock Principles of parliamentary representation.
\newblock 1884.

\bibitem{john1971rawls}
John~W. Chapman.
\newblock Rawls's theory of justice.
\newblock 1971.

\bibitem{chen2021decentralized}
Yan Chen, Jack~I Richter, and Pankaj~C Patel.
\newblock Decentralized governance of digital platforms.
\newblock {\em Journal of Management}, 47(5), 2021.

\bibitem{swarm-city}
Swarm City.
\newblock Swarm city.
\newblock \url{https://swarm.city}, 2023.
\newblock Accessed on February 2, 2023.

\bibitem{Compound@CoinGecko}
{CoinGecko}.
\newblock {Compound Tokenomics}.
\newblock \url{https://www.coingecko.com/en/coins/compound/tokenomics}, 2023.
\newblock Accessed on February 2, 2023.

\bibitem{Uniswap@CoinGecko}
{CoinGecko}.
\newblock {Uniswap Tokenomics}.
\newblock \url{https://www.coingecko.com/en/coins/uniswap/tokenomics}, 2023.
\newblock Accessed on February 2, 2023.

\bibitem{Governance@Compound}
{Compound Labs, Inc.}
\newblock {Compound Governance}.
\newblock \url{https://docs.compound.finance/governance}, 2022.
\newblock Accessed on Dec 10, 2022.

\bibitem{Governance@ConvexFinance}
{Convex}.
\newblock {Convex Finance Proposals}.
\newblock \url{https://vote.convexfinance.com/}, 2023.
\newblock Accessed on May 25, 2023.

\bibitem{Daian@S&P20}
Philip Daian, Steven Goldfeder, Tyler Kell, Yunqi Li, Xueyuan Zhao, Iddo Bentov, Lorenz Breidenbach, and Ari Juels.
\newblock Flash boys 2.0: Frontrunning in decentralized exchanges, miner extractable value, and consensus instability.
\newblock In {\em 2020 IEEE Symposium on Security and Privacy (SP)}, 2020.

\bibitem{Daian-dark-dao@HackingDistributed}
Philip Daian, Tyler Kell, Ian Miers, and Ari Juels.
\newblock {On-Chain Vote Buying and the Rise of Dark DAOs}.
\newblock \url{https://hackingdistributed.com/2018/07/02/on-chain-vote-buying}, 2018.
\newblock Accessed on December 15, 2022.

\bibitem{dotan2023vulnerable}
Maya Dotan, Aviv Yaish, Hsin-Chu Yin, Eytan Tsytkin, and Aviv Zohar.
\newblock The vulnerable nature of decentralized governance in defi.
\newblock In {\em Proceedings of the 2023 Workshop on Decentralized Finance and Security}, DeFi '23, page 25–31, 2023.

\bibitem{SVB@WallStreet}
{Edward Helmore}.
\newblock {Why did the \$212bn tech-lender Silicon Valley bank abruptly collapse?}
\newblock \url{https://www.theguardian.com/business/2023/mar/17/why-silicon-valley-bank-collapsed-svb-fail}, 2023.
\newblock Accessed on April 10, 2023.

\bibitem{NFTs}
{Ethereum Foundation}.
\newblock {Non-fungible tokens (NFT)}.
\newblock \url{https://ethereum.org/en/nft}, 2023.
\newblock Accessed on May 25, 2023.

\bibitem{Etherscan@ETH-explorer}
{Etherscan}.
\newblock {Etherscan (ETH) Blockchain Explorer}.
\newblock \url{https://etherscan.io}, 2023.
\newblock Accessed on May 25, 2023.

\bibitem{andres2025dao}
Andr{\'e}s F{\'a}brega, Amy Zhao, Jay Yu, James Austgen, Sarah Allen, Kushal Babel, Mahimna Kelkar, and Ari Juels.
\newblock Voting-bloc entropy: A new metric for dao decentralization.
\newblock In {\em Proceedings of the 34th USENIX Security Symposium (USENIX Security 25)}, 2025.

\bibitem{feichtinger2024sok}
Rainer Feichtinger, Robin Fritsch, Lioba Heimbach, Yann Vonlanthen, and Roger Wattenhofer.
\newblock Sok: Attacks on daos.
\newblock In {\em Proceedings of the 6th ACM Conference on Advances in Financial Technologies}, AFT '24, 2024.

\bibitem{feichtinger2023hidden}
Rainer Feichtinger, Robin Fritsch, Yann Vonlanthen, and Roger Wattenhofer.
\newblock The hidden shortcomings of (d)aos -- an empirical study of on-chain governance, 2024.

\bibitem{fritsch@2022votingpower}
Robin Fritsch, Marino Müller, and Roger Wattenhofer.
\newblock Analyzing voting power in decentralized governance: Who controls daos?, 2024.
\newblock \href {https://doi.org/10.1016/j.bcra.2024.100208} {\path{doi:10.1016/j.bcra.2024.100208}}.

\bibitem{Gauntlet@Compound}
Gauntlet.
\newblock {Gauntlet <> Compound Renewal}.
\newblock \url{https://www.comp.xyz/t/gauntlet-compound-renewal/3541}, 2022.
\newblock Accessed on April 7, 2025.

\bibitem{hassan2021decentralized}
Samer Hassan and Primavera De~Filippi.
\newblock Decentralized autonomous organization.
\newblock {\em Internet Policy Review}, 10(2), 2021.

\bibitem{hobbesleviathan}
Thomas Hobbes.
\newblock Leviathan.
\newblock {\em Project Gutenberg}, 1651.

\bibitem{Alameda@CoinDesk}
{Ian Allison}.
\newblock {Divisions in Sam Bankman-Fried’s Crypto Empire Blur on His Trading Titan Alameda’s Balance Sheet}.
\newblock \url{https://www.coindesk.com/business/2022/11/02/divisions-in-sam-bankman-frieds-crypto-empire-blur-on-his-trading-titan-alamedas-balance-sheet/}, 2022.
\newblock Accessed on Dec 10, 2022.

\bibitem{kiayias@2022governance}
Aggelos Kiayias and Philip Lazos.
\newblock Sok: Blockchain governance, 2023.
\newblock \href {https://doi.org/10.1145/3558535.3559794} {\path{doi:10.1145/3558535.3559794}}.

\bibitem{kitzler2023governance}
Stefan Kitzler, Stefano Balietti, Pietro Saggese, Bernhard Haslhofer, and Markus Strohmaier.
\newblock The governance of decentralized autonomous organizations: A study of contributors' influence, networks, and shifts in voting power, 2025.

\bibitem{Luna@CoinDesk}
{Krisztian Sandor and Ekin Genç}.
\newblock {The Fall of Terra: A Timeline of the Meteoric Rise and Crash of UST and LUNA}.
\newblock \url{https://www.coindesk.com/learn/the-fall-of-terra-a-timeline-of-the-meteoric-rise-and-crash-of-ust-and-luna/}, 2022.
\newblock Accessed on February 2, 2023.

\bibitem{kybx86@Compound}
kybx86.
\newblock Compensation proposal: Distribute comp to affected users in the dai liquidations.
\newblock \url{https://www.comp.xyz/t/compensation-proposal-distribute-comp-to-affected-users-in-the-dai-liquidations}, 2020.
\newblock Accessed on February 9, 2023.

\bibitem{Snapshot}
Snapshot Labs.
\newblock Snapshot.
\newblock \url{https://snapshot.org}, 2023.
\newblock Accessed on February 2, 2023.

\bibitem{Uniswap-Proposals-Page}
Uniswap Labs.
\newblock Uniswp proposals.
\newblock \url{https://app.uniswap.org/vote}, 2023.
\newblock Accessed on September 20, 2023.

\bibitem{Lalley@AEA}
Steven~P. Lalley and E.~Glen Weyl.
\newblock Quadratic voting: How mechanism design can radicalize democracy.
\newblock {\em AEA Papers and Proceedings}, 108, May 2018.

\bibitem{leshner2019compound}
Robert Leshner and Geoffrey Hayes.
\newblock Compound: The money market protocol, 2019.

\bibitem{Ma@ACM-Transactions}
Wei Ma, Chenguang Zhu, Ye~Liu, Xiaofei Xie, and Yi~Li.
\newblock A comprehensive study of governance issues in decentralized finance applications.
\newblock {\em ACM Trans. Softw. Eng. Methodol.}, February 2025.
\newblock \href {https://doi.org/10.1145/3717062} {\path{doi:10.1145/3717062}}.

\bibitem{Governance@MakerDAO}
{MakerDAO}.
\newblock {Governance Module -- Maker Protocol Technical Docs}.
\newblock \url{https://docs.makerdao.com/smart-contract-modules/governance-module}, 2023.
\newblock Accessed on April 2, 2023.

\bibitem{Mastodon}
{Mastodon}.
\newblock \url{https://joinmastodon.org}, 2024.
\newblock Accessed on May 15, 2024.

\bibitem{Messias@IMC2021}
Johnnatan Messias, Mohamed Alzayat, Balakrishnan Chandrasekaran, Krishna~P. Gummadi, Patrick Loiseau, and Alan Mislove.
\newblock {Selfish \& Opaque Transaction Ordering in the Bitcoin Blockchain: The Case for Chain Neutrality}.
\newblock In {\em Proceedings of the ACM Internet Measurement Conference (IMC'21)}, November 2021.

\bibitem{Messias-Dataset-Code-2025}
Johnnatan Messias, Vabuk Pahari, Balakrishnan Chandrasekaran, Krishna~P. Gummadi, and Patrick Loiseau.
\newblock {Data set and scripts used for analyzing the power dynamics in blockchain governance}.
\newblock \url{https://github.com/johnnatan-messias/research-proj-governance}, 2025.

\bibitem{Motepalli@ICBC}
Shashank Motepalli and Hans-Arno Jacobsen.
\newblock How does stake distribution influence consensus? analyzing blockchain decentralization.
\newblock In {\em 2024 IEEE International Conference on Blockchain and Cryptocurrency (ICBC)}, pages 343--352, 2024.
\newblock \href {https://doi.org/10.1109/ICBC59979.2024.10634400} {\path{doi:10.1109/ICBC59979.2024.10634400}}.

\bibitem{Nakamoto-WhitePaper2008}
Satoshi Nakamoto.
\newblock {Bitcoin: A Peer-to-Peer Electronic Cash System}, 2008.

\bibitem{Alameda@Investopedia}
{Nathan Reiff}.
\newblock {The Collapse of FTX: What Went Wrong with the Crypto Exchange?}
\newblock \url{https://www.investopedia.com/what-went-wrong-with-ftx-6828447}, 2023.
\newblock Accessed on April 2, 2023.

\bibitem{Perez@FC21}
Daniel Perez, Sam~M Werner, Jiahua Xu, and Benjamin Livshits.
\newblock Liquidations: Defi on a knife-edge.
\newblock In {\em Financial Cryptography and Data Security}, FC '21, 2021.

\bibitem{Luna@Forbes}
{Q.ai}.
\newblock {What Really Happened To LUNA Crypto?}
\newblock \url{https://www.forbes.com/sites/qai/2022/09/20/what-really-happened-to-luna-crypto/?sh=62332b664ff1}, 2022.
\newblock Accessed on December 10, 2022.

\bibitem{Qin@FC21}
Kaihua Qin, Liyi Zhou, Benjamin Livshits, and Arthur Gervais.
\newblock {Attacking the DeFi Ecosystem with Flash Loans for Fun and Profit}.
\newblock In {\em Financial Cryptography and Data Security}, FC '21, 2021.

\bibitem{raman2019challenges}
Aravindh Raman, Sagar Joglekar, Emiliano~De Cristofaro, Nishanth Sastry, and Gareth Tyson.
\newblock Challenges in the decentralised web: The mastodon case, 2019.
\newblock \href {https://doi.org/10.1145/3355369.3355572} {\path{doi:10.1145/3355369.3355572}}.

\bibitem{rikken2019governance}
Olivier Rikken, Marijn Janssen, and Zenlin Kwee.
\newblock Governance challenges of blockchain and decentralized autonomous organizations.
\newblock {\em Information Polity}, 24(4), 2019.

\bibitem{rousseau1920social}
Jean-Jacques Rousseau.
\newblock {\em The social contract: \& discourses}.
\newblock Number 660. JM Dent \& Sons, 1920.

\bibitem{sasson2014zerocash}
Eli~Ben Sasson, Alessandro Chiesa, Christina Garman, Matthew Green, Ian Miers, Eran Tromer, and Madars Virza.
\newblock Zerocash: Decentralized anonymous payments from bitcoin.
\newblock In {\em 2014 IEEE symposium on security and privacy}. IEEE, 2014.

\bibitem{Cosine@ScikitLearn}
{Scikit Learn}.
\newblock {Cosine Similarity}.
\newblock \url{https://scikit-learn.org/stable/modules/generated/sklearn.metrics.pairwise.cosine_similarity.html}, 2023.
\newblock Accessed on April 10, 2023.

\bibitem{sharma2023unpacking}
Tanusree Sharma, Yujin Potter, Kornrapat Pongmala, Henry Wang, Andrew Miller, Dawn Song, and Yang Wang.
\newblock Unpacking how decentralized autonomous organizations (daos) work in practice.
\newblock In {\em 2024 IEEE International Conference on Blockchain and Cryptocurrency (ICBC)}, 2024.

\bibitem{Binance-denies@coindesk}
{Shaurya Malwa}.
\newblock {Binance Denies Allegations It Intends to Use Users' Uniswap Tokens for Voting}.
\newblock \url{https://www.coindesk.com/tech/2022/10/20/binance-denies-allegations-that-it-intends-to-use-users-uniswap-tokens-for-voting/}, 2022.

\bibitem{Addresses@Sybil}
{Sybil}.
\newblock {Sybil -- Top delegated addresses}.
\newblock \url{https://sybil.org/}, 2023.
\newblock Accessed on February 2, 2023.

\bibitem{Tally}
{Tally}.
\newblock {Tally: Run DAOs onchain}.
\newblock \url{https://www.tally.xyz}, 2024.
\newblock Accessed on September 2, 2024.

\bibitem{Arbitrum@Tally}
Tally.
\newblock {Arbitrum Delegates}.
\newblock \url{https://www.tally.xyz/gov/arbitrum/delegates}, 2025.
\newblock Accessed on February 24, 2025.

\bibitem{Compound@Tally}
Tally.
\newblock {Compound Delegates}.
\newblock \url{https://www.tally.xyz/gov/compound/delegates}, 2025.
\newblock Accessed on February 24, 2025.

\bibitem{ENS@Tally}
Tally.
\newblock {ENS Delegates}.
\newblock \url{https://www.tally.xyz/gov/ens/delegates}, 2025.
\newblock Accessed on February 24, 2025.

\bibitem{Uniswap@Tally}
Tally.
\newblock {Uniswap Delegates}.
\newblock \url{https://www.tally.xyz/gov/uniswap/delegates}, 2025.
\newblock Accessed on February 24, 2025.

\bibitem{Wormhole@Tally}
Tally.
\newblock {Wormhole Delegates}.
\newblock \url{https://www.tally.xyz/gov/wormhole/delegates}, 2025.
\newblock Accessed on February 24, 2025.

\bibitem{ZKsync@Tally}
Tally.
\newblock {ZKsync Delegates}.
\newblock \url{https://www.tally.xyz/gov/zksync/delegates}, 2025.
\newblock Accessed on February 24, 2025.

\bibitem{tan2023open}
Joshua~Z Tan, Tara Merk, Sarah Hubbard, Eliza~R Oak, Joni Pirovich, Ellie Rennie, Rolf Hoefer, Michael Zargham, Jason Potts, Chris Berg, et~al.
\newblock Open problems in daos.
\newblock {\em arXiv preprint arXiv:2310.19201}, 2023.

\bibitem{Uniswap-Lifecycle}
{Uniswap Labs}.
\newblock {Uniswap: Governance Proposal Process}.
\newblock \url{https://docs.uniswap.org/concepts/governance/process}, 2024.
\newblock Accessed on October 11, 2024.

\bibitem{van2013cryptonote}
Nicolas Van~Saberhagen.
\newblock Cryptonote v2.0.
\newblock 2013.

\bibitem{Weintraub@IMC2022}
Ben Weintraub, Christof~Ferreira Torres, Cristina Nita-Rotaru, and Radu State.
\newblock {A Flash(bot) in the Pan: Measuring Maximal Extractable Value in Private Pools}.
\newblock In {\em Proceedings of the ACM Internet Measurement Conference (IMC'22)}, October 2022.

\bibitem{Rich-gets-Richer@Wikimedia}
Inc Wikimedia~Foundation.
\newblock {The rich get richer and the poor get poorer}.
\newblock \url{https://en.wikipedia.org/wiki/The_rich_get_richer_and_the_poor_get_poorer}, 2025.
\newblock Accessed on February 24, 2025.

\bibitem{Wood@Ethereum}
Gavin Wood.
\newblock Ethereum: A secure decentralised generalised transaction ledger, 2014.

\bibitem{wu2020coefficient}
Keke Wu, Bo~Peng, Hua Xie, and Shaobin Zhan.
\newblock A coefficient of variation method to measure the extents of decentralization for bitcoin and ethereum networks.
\newblock {\em Int. J. Netw. Secur.}, 22(2):191--200, 2020.

\bibitem{xia2015learning}
Peipei Xia, Li~Zhang, and Fanzhang Li.
\newblock Learning similarity with cosine similarity ensemble.
\newblock {\em Information sciences}, 307, 2015.

\bibitem{eth-usd@yahoo}
{Yahoo Finance}.
\newblock {Ethereum USD (ETH-USD), Price, Value, News, and History}.
\newblock \url{https://finance.yahoo.com/quote/ETH-USD/history?p=ETH-USD}, 2023.
\newblock Accessed on August 20, 2024.

\bibitem{yaish2024strategic}
Aviv Yaish, Svetlana Abramova, and Rainer B{\"o}hme.
\newblock Strategic vote timing in online elections with public tallies.
\newblock {\em arXiv preprint arXiv:2402.09776}, 2024.

\bibitem{zwitter2020decentralized}
Andrej Zwitter and Jilles Hazenberg.
\newblock Decentralized network governance: blockchain technology and the future of regulation.
\newblock {\em Frontiers in Blockchain}, 3, 2020.

\end{thebibliography}

\appendix

\section{Time Until Reaching the Quorum in Compound}\label{sec:compound-quorum}

\begin{figure*}[tb]
\centering
\includegraphics[width=\linewidth]{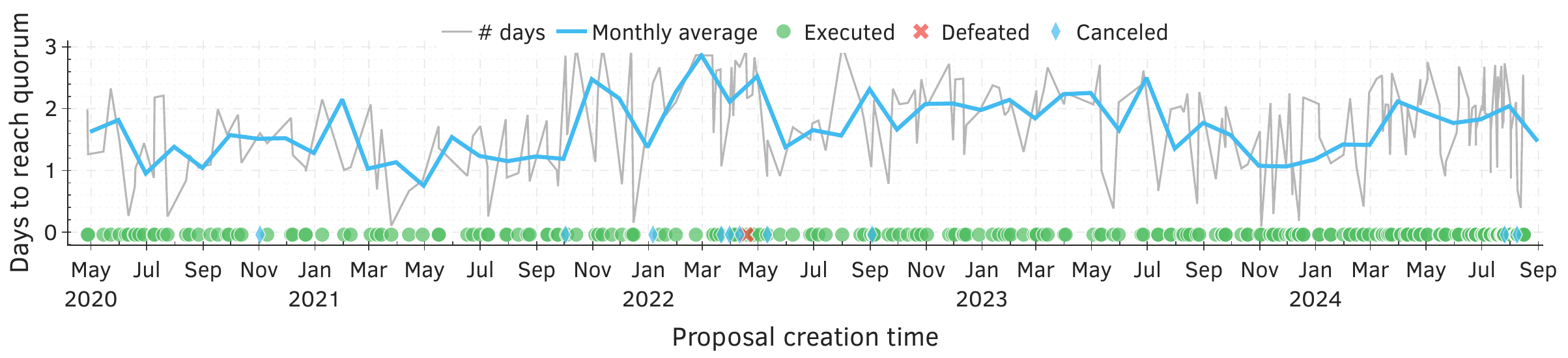}
\caption{Compound proposals typically reach the quorum after 1.69 days on average (with a standard deviation of 0.69 and a median of 1.81 days.}
\label{fig:new_compound-time-to-quorum}
\end{figure*}

\begin{figure*}[tb]
\centering
\includegraphics[width=\linewidth]{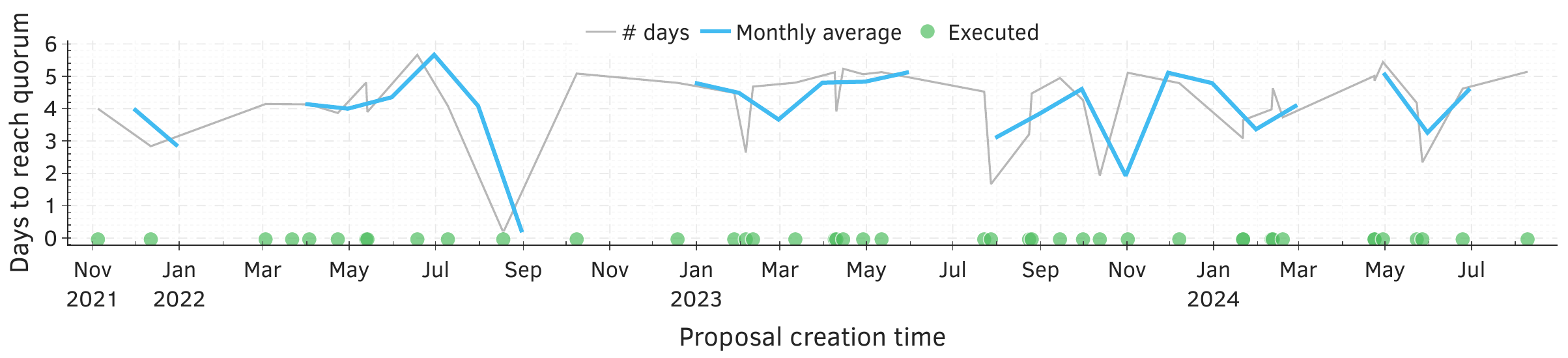}
\caption{Uniswap proposals typically reach the quorum after 4.15 days on average (with a standard deviation of 1.11 and a median of 4.47 days.}
\label{fig:new_uniswap-time-to-quorum}
\end{figure*}

Typically, for a governance proposal to pass, it must receive a majority of votes in favor and reach a specified quorum. In both Compound and Uniswap, the quorum is set as \num{4}\% of the total supply, accounting for \num{400000} and \num{40000000}, respectively.
This minimum number of in favor votes is referred to as \stress{quorum} and is defined by the Governor Bravo contract.

We analyze how long it takes for these proposals to reach the required quorum.
Figure~\ref{fig:new_compound-time-to-quorum} shows the number of days it took each of the Compound proposals evaluated to reach the quorum, and Figure~\ref{fig:new_uniswap-time-to-quorum} for Uniswap.
On average, Compound proposals take \num{1.69} days with a std. of \num{0.69} days to reach the quorum while for Uniswap, it takes \num{4.15} days with a std. of \num{1.11} days.

\end{document}